\documentclass[12pt]{article}
\pdfoutput=1

\usepackage[utf8]{inputenc}
\usepackage[left=2.55cm, right=2.55cm, top=2.55cm, bottom=2.55cm]{geometry}
\usepackage{amsmath,amssymb,amsbsy}
\usepackage{slashed}
\usepackage{xcolor} 
\usepackage{graphicx}
\usepackage{url}
\usepackage{cancel}
\usepackage{cite}
\usepackage[colorlinks=true,allcolors=darkpurple,pdfborder={0 0 0},linktocpage=false,pdfencoding=auto]{hyperref}
\usepackage{tabularx,booktabs}
\usepackage{multicol}
\usepackage{feynmp}
\usepackage{units}
\usepackage{xspace}
\usepackage[labelfont=bf]{caption}
\usepackage[section]{placeins}
\usepackage{subcaption}
\usepackage{soul} 

\definecolor{darkred}{rgb}{0.6,0,0}
\definecolor{darkpurple}{rgb}{0.5,0,0.5}

\def\hc{\text{h.c.}}

\def\z2{$\mathbb{Z}_2$}
\def\321{$\mathrm{SU(3)_c} \times \mathrm{SU(2)_L} \times \mathrm{U(1)_Y}$}
\def\one{\ensuremath{\mathbf{1}}}
\def\two{\ensuremath{\mathbf{2}}}
\def\three{\ensuremath{\mathbf{3}}}
\def\threeS{\ensuremath{\mathbf{\bar 3}}}

\newcommand{\AddrIFIC}{%
  Instituto de F\'{i}sica Corpuscular, CSIC-Universitat de Val\`{e}ncia, 46980 Paterna, Spain}

\newcommand{\AddrFISTEO}{%
  Departament de F\'{\i}sica Te\`{o}rica, Universitat de Val\`{e}ncia, 46100 Burjassot, Spain}

\newcommand{\AddrIAC}{%
  Instituto de Astrof\'isica de Canarias, C/ V\'ia L\'actea, s/n, 38205 La Laguna, Tenerife, Spain}

 \newcommand{\AddrULL}{
 Universidad de La Laguna, Departamento de Astrof\'isica, La Laguna, Tenerife, Spain
 }%

\begin{document}

\vspace*{-2cm}
\begin{flushright}
IFIC/21-07 \\
\vspace*{2mm}
\end{flushright}

\begin{center}
\vspace*{15mm}

\vspace{1cm} {\Large \bf
  $\boldsymbol{(g-2)_{e,\mu}}$ in an extended inverse type-III seesaw
} \\
\vspace{1cm}

{\bf Pablo Escribano$^{\text{a}}$, Jorge Terol-Calvo$^{\text{b,c}}$, Avelino Vicente$^{\text{a,d}}$}

 \vspace*{.5cm} 
 $^{(\text{a})}$ \AddrIFIC \\\vspace*{.2cm} 
 $^{(\text{b})}$ \AddrIAC \\\vspace*{.2cm}  
 $^{(\text{c})}$ \AddrULL \\\vspace*{.2cm} 
 $^{(\text{d})}$ \AddrFISTEO

 \vspace*{.3cm} 
\href{mailto:pablo.escribano@ific.uv.es}{pablo.escribano@ific.uv.es}, \href{mailto:jorgetc@iac.es}{jorgetc@iac.es}, \href{mailto:avelino.vicente@ific.uv.es}{avelino.vicente@ific.uv.es}
\end{center}

\vspace*{10mm}
\begin{abstract}\noindent\normalsize
There has been a long-standing discrepancy between the experimental measurements of the electron and muon anomalous magnetic moments and their predicted values in the Standard Model. This is particularly relevant in the case of the muon $g-2$, which has attracted a remarkable interest in the community after the long-awaited announcement of the first results by the Muon $g-2$ collaboration at Fermilab, which confirms a previous measurement by the E821 experiment at Brookhaven and enlarges the statistical significance of the discrepancy, now at $4.2 \sigma$. In this paper we consider an extension of the inverse type-III seesaw with a pair of vector-like leptons that induces masses for neutrinos at the electroweak scale and show that one can accommodate the electron and muon anomalous magnetic moments, while being compatible with all relevant experimental constraints.
\end{abstract}




\section{Introduction}
\label{sec:intro}

The charged leptons anomalous magnetic moments,
\begin{equation}
  a_\ell = \frac{g_\ell - 2}{2} \, ,
\end{equation}
with $\ell = e, \mu, \tau$, are known to be powerful probes of New
Physics (NP) effects, potentially hidden in virtual loop
contributions. Interestingly, there has been a long-standing
discrepancy between the Standard Model (SM) prediction for the
electron and muon anomalous magnetic moments and their experimentally
determined
values~\cite{Aoyama:2012wj,Aoyama:2012wk,Laporta:2017okg,Aoyama:2017uqe,Bennett:2006fi,Jegerlehner:2009ry,Blum:2018mom}. In
the case of the electron $g-2$, the significance is slightly below
$\sim 3 \sigma$, and hence not very significant at the moment. In
contrast, the deviation has become particularly relevant in the case
of the muon $g-2$, in particular after the Muon $g-2$ experiment at
Fermilab has published its long-awaited first
results~\cite{PhysRevLett.126.141801}. Their measurement of $a_\mu$
perfectly agrees with the result obtained by the E821 experiment at
Brookhaven~\cite{Bennett:2006fi} and, consequently, disagrees with the
SM. Their combination leads to a $4.2 \sigma$ discrepancy with the SM
prediction compiled by the theory community
in~\cite{Aoyama:2020ynm}. In summary, the current status of the
electron and muon $g-2$ can be quantified as~\footnote{The status of
the electron $g-2$ has recently changed by a new measurement of the
fine-structure constant~\cite{Morel:2020dww}. The new value differs by
more than $5 \sigma$ to the previous one and affects the electron
$g-2$ anomaly, which gets reduced to just $1.6 \sigma$ and flips sign,
see~\cite{Gerardin:2020gpp}. We will not include these results in our
analysis, but note that it would be straightforward to accommodate a
positive $\Delta a_e$ in our model, as shown in
Sec.~\ref{sec:pheno}. We also point out that this change in the
fine-structure constant value has little impact on the muon $g-2$.}
\begin{align} 
  \Delta a_e &= a_e^{\text{exp}} - a_e^{\text{SM}} = (-87 \pm 36) \times 10^{-14} \, , \nonumber \\
  \Delta a_{\mu} &= a_\mu^{\text{exp}} - a_\mu^{\text{SM}} = (25.1 \pm 5.9) \times 10^{-10} \, . \label{eq:Deltaa}
\end{align}

New measurements and more refined theoretical calculations are
definitely required to assess the relevance of these anomalies, and
confirm whether these intriguing deviations are hints of
NP~\cite{Lindner:2016bgg}, SM contributions not correctly taken into
account or just statistical fluctuations.~\footnote{The theoretical
  calculation of the electron and muon anomalous magnetic moments is a
  challenging task and has led to some controversies along the
  years. For instance, a recent calculation of the hadronic vacuum
  polarization contribution by the Budapest-Marseilles-Wuppertal
  collaboration~\cite{Borsanyi:2020mff} brings the SM prediction for
  the muon $g-2$ into agreement with the experimental value, hence
  ruling out any discrepancy. However, it has been pointed out that
  this result in turn leads to some tension with electroweak
  data~\cite{Passera:2008jk,Crivellin:2020zul,Keshavarzi:2020bfy,Malaescu:2020zuc}.}
However, it is tempting to interpret them as a signal of the presence
of new states beyond the SM (BSM). In this case, the $g-2$ anomalies
may hide valuable information about the shape of the underlying
model. In particular, the sign difference between $\Delta a_e$ and
$\Delta a_\mu$ and the sizable value of $|\Delta a_e|$ would indicate
that the NP contributions do not scale with the square of the charged
lepton masses~\cite{Giudice:2012ms}. This calls for a non-trivial
extension of the SM.

The inverse type-III seesaw model (ISS3) is obtained by replacing the
fermionic $\mathrm{SU(2)_L}$ singlets in the original inverse type-I
seesaw~\cite{Mohapatra:1986bd} by $\mathrm{SU(2)_L}$ triplets. This
variant has already been
studied~\cite{Abada:2007ux,Gavela:2009cd,Ibanez:2009du,Ma:2009kh,Eboli:2011ia,Morisi:2012hu,Aguilar-Saavedra:2013twa},
although not extensively, and many of its phenomenological features
are still to be investigated.~\footnote{See
  also~\cite{Law:2013gma,CentellesChulia:2020dfh} for discussions of
  generalized inverse seesaw models, including versions with Dirac
  neutrinos,
  \cite{Abada:2008ea,Franceschini:2008pz,Das:2020uer,Ashanujjaman:2021jhi}
  for four references studying the phenomenology of light fermion
  triplets and \cite{Mandal:2021acg} for a recent work on the inverse
  seesaw with spontaneous violation of lepton number.} In fact, there
are several phenomenological directions of interest in the ISS3. The
fact that triplets couple to the SM gauge bosons allow for new
production mechanisms at the LHC, where one can also look for lepton
number violating signatures. Lepton flavor violation might also be an
interesting subject to explore in this model, which may offer some
differences with respect to the more common inverse type-I
seesaw~\cite{Abada:2014kba}. Finally, the potentially sizable mixing
between the charged components of the type-III triplet and the SM
charged leptons may also lead to observable $Z \to \ell^+_i \ell^-_j$
decays with $i \ne j$. We refer to~\cite{Biggio:2019eeo} for a recent
analysis of these and other relevant observables in the presence of
light fermion triplets.

While many models have been put forward to address the discrepancy
between theory and experiment in the electron and muon $g-2$, the main
motivation for the ISS3 is not to accommodate the existing deviations,
but to induce non-zero masses for neutrinos. It is therefore natural
to investigate whether the model can account for the experimental
values for the electron and muon anomalous magnetic moments in the
region of parameter space that can reproduce the observed neutrino
masses and leptonic mixing angles, measured in oscillation
experiments, while being compatible with the bounds obtained at
colliders and low-energy experiments. In this paper we show that these
constraints preclude the ISS3 from inducing large contributions to
$(g-2)_{e,\mu}$. More importantly, the ISS3 contributions are
negative, making it impossible to address the existing discrepancy in
the muon $g-2$. This motivates a minimal extension of the model
that keeps its most relevant features but provides additional
ingredients to generate the required contributions to the electron and
muon $g-2$. We find that the introduction of a pair of vector-like
(VL) lepton doublets with sizable couplings to electrons and muons can
explain both anomalies and simultaneously satisfy all the experimental
constraints. It is the aim of this paper to study the electron and
muon $g-2$ in this model, which we denote as the ISS3VL.

The muon $g-2$ has been considered in a wide variety of contexts, in
many cases in connection to neutrino mass generation. This includes
models based on the inverse seesaw
mechanism~\cite{Abdallah:2011ew,Khalil:2015wua,Cao:2019evo,Dinh:2020pqn,Cao:2021lmj,Nomura:2021adf,Mondal:2021vou,Hue:2021xap,CarcamoHernandez:2021iat}
and/or with VL
leptons~\cite{Dermisek:2013gta,Poh:2017tfo,Kowalska:2017iqv,Megias:2017dzd,Chiang:2017tai,Calibbi:2018rzv,Arnan:2019uhr,Kawamura:2019rth,Calibbi:2020emz,Frank:2020smf,Chun:2020uzw,Chakrabarty:2020jro,Chen:2020tfr,Jana:2020joi,Dermisek:2020cod,Das:2020hpd,Baker:2021yli,Dermisek:2021ajd,Das:2021zea,Chiang:2021pma,Arcadi:2021cwg}. See
also~\cite{Arbelaez:2020rbq} for a recent work in the context of a
radiative neutrino mass model including triplet fermions. Finally, we
note that the muon $g-2$ has also been considered as a motivation for
a muon
collider~\cite{Buttazzo:2020eyl,Yin:2020afe,Capdevilla:2021rwo}.

The rest of the manuscript is organized as follows. In
Sec.~\ref{sec:model} the basic features of the ISS3VL model are
introduced, including the generation of charged and neutral lepton
masses. In Sec.~\ref{sec:g-2} we compute the charged lepton $g-2$
values and provide simplified approximate expressions. A numerical
analysis is performed in Sec.~\ref{sec:pheno}. After arguing that the
pure ISS3 model cannot address the anomalies, we explore the parameter
space of the ISS3VL model and obtain results for the electron and muon
$g-2$ compatible with the relevant experimental constraints. Finally,
we discuss our results and conclude in
Sec.~\ref{sec:conclusions}. Appendices~\ref{sec:app1} and
\ref{sec:app2} contain additional details, such as analytical
expressions for the couplings of interest to our calculation and full
expressions for the charged lepton anomalous magnetic moments.

\section{The model}
\label{sec:model}

The ISS3VL is an extension of the leptonic sector of the SM with the
addition of six right-handed Weyl fermion $\mathrm{SU(2)_L}$ triplets
with vanishing hypercharge, $\Sigma_{A}$ and $\Sigma_{A}^\prime$
($A=1,2,3$), and a VL copy of the SM lepton doublet, $L_L$ and
$L_R$. The $\Sigma_{A}$ and $\Sigma_{A}^\prime$ triplets are
introduced in order to generate neutrino masses via the inverse
type-III seesaw mechanism.~\footnote{In order to simplify the
  notation, we will not denote the chirality of the $\Sigma_{A} \equiv
  \Sigma_{R_A}$ and $\Sigma_{A}^\prime \equiv \Sigma_{R_A}^\prime$
  fermions explicitly.} They can be distinguished by their different
lepton numbers, with $L(\Sigma)=+1$ and
$L(\Sigma^\prime)=-1$. Nevertheless, lepton number will be explicitly
broken in the ISS3VL. Therefore, this lepton number assignment is
arbitrary. The new fermionic fields $L_L$ and $L_R$ have the same
representations under the \321 gauge group, and both are doublets
under $\mathrm{SU(2)_L}$. The full particle content of the ISS3VL
model and the representations of all fields under the \321 gauge group
are shown in Tab.~\ref{tab:content}.

As usual, the SM $\mathrm{SU(2)_L}$ doublets can be decomposed as
\begin{equation}
q_L = \left( \begin{array}{c}
u \\
d
\end{array} \right)_L \quad , \quad \ell_L = \left( \begin{array}{c}
\nu \\
e
\end{array} \right)_L \quad , \quad H = \left( \begin{array}{c}
H^+ \\
H^0
\end{array} \right) \, .
\end{equation}
The $\Sigma$ and $\Sigma^\prime$ triplets can also be decomposed into
$\rm SU(2)_L$ components. With
$\Sigma_{A}=(\Sigma^1,\Sigma^2,\Sigma^3)_A$, they can be conveniently
written in the usual $2 \times 2$ matrix notation according to (the
same holds for the primed states)
\begin{equation}
  \label{eq:triplet-2times2}
  \Sigma_{A}= \frac{1}{\sqrt{2}} \, \vec{\tau} \cdot \vec{\Sigma}_{A}=
  \begin{pmatrix}
    \Sigma_{A}^0/\sqrt{2} & \Sigma_{A}^+\\
    \Sigma_{A}^- & -\Sigma_{A}^0/\sqrt{2} 
  \end{pmatrix}\, ,
\end{equation}
where $\tau_A$ are the usual Pauli matrices and the states with
well-defined electric charge are given by  
\begin{equation}
  \label{eq:2times2-triplet}
  \Sigma_{A}^0=\Sigma_{A}^3\, ,\qquad
  \Sigma_{A}^\pm =
  \frac{\Sigma_{A}^1\mp i \, \Sigma_{A}^2}{\sqrt{2}}\, .
\end{equation}
Finally, the VL leptons $L_{L,R}$ can decomposed as
\begin{equation}
L_{L,R} = \left( \begin{array}{c}
N \\
E
\end{array} \right)_{L,R} \, .
\end{equation}

{
\renewcommand{\arraystretch}{1.6}
\begin{table}[tb]
\centering
\begin{tabular}{ c | c c c c c c c c c | c }
\toprule
& $q_L$ & $u_R$ & $d_R$ & $\ell_L$ & $e_R$ & $\Sigma$ & $\Sigma^\prime$ & $L_L$ & $L_R$ & $H$\\ 
\hline
$\rm SU(3)_C$ & $\three$ & $\threeS$ & $\threeS$ & $\one$ & $\one$ & $\one$ & $\one$ & $\one$ & $\one$ &$\one$ \\
$\rm SU(2)_L$ & $\two$ & $\one$ & $\one$ & $\two$ & $\one$ & $\three$ & $\three$ & $\two$ & $\two$ & $\two$ \\
$\rm U(1)_Y$ & $\frac{1}{6}$ & $\frac{2}{3}$ & $-\frac{1}{3}$ & $-\frac{1}{2}$ & $-1$ & $0$ & $0$ & $-\frac{1}{2}$ & $-\frac{1}{2}$ & $\frac{1}{2}$ \\[1mm]
\hline
\textsc{Generations} & 3 & 3 & 3 & 3 & 3 & 3 & 3 & 1 & 1 & 1 \\
\bottomrule
\end{tabular}
\caption{Particle content of the ISS3VL. $q_L$, $\ell_L$, $u_R$, $d_R$,
  $e_R$ and $H$ are the usual SM fields. 
\label{tab:content}}
\end{table}
}

Under the above working assumptions, the most general Yukawa
Lagrangian allowed by all symmetries can be written as
\begin{equation} \label{eq:lag}
\mathcal{L}_Y = \mathcal{L}_Y^{\text{SM}} + \mathcal{L}_Y^{\text{ISS3}} + \mathcal{L}_Y^{\text{VL}} \, ,
\end{equation}
where
\begin{equation} \label{eq:YukawaSM}
-\mathcal{L}_Y^{\text{SM}} =
  \overline{q}_L \, Y_u \, u_R \, \widetilde{H}
  + \overline{q}_L \, Y_d \, d_R \, H
  + \overline{\ell}_L \, Y_e \, e_R \, H
  + \hc
\end{equation}
is the usual SM Lagrangian, with $\widetilde{H}=i\,\tau_2\,H^*$, and
$Y_{u,d,e}$ are $3\times 3$ Yukawa matrices in flavor space. Here and
in the following we omit $\mathrm{SU(2)_L}$ contractions and flavor
indices to simplify the notation. The terms
\begin{equation} \label{eq:ISS3}
-\mathcal{L}_Y^{\text{ISS3}} = 
\sqrt{2} \, \overline{\Sigma}\,Y_\Sigma\,\ell_L\,\widetilde{H}^\dagger
+ \overline{\Sigma} \, M_\Sigma \, \Sigma^{\prime c} 
+ \frac{1}{2} \, \overline{\Sigma^\prime}\,\mu\,\Sigma^{\prime c}
+ \, \hc \, ,
\end{equation}
correspond to the usual ISS3 extension. Here $Y_\Sigma$, $M_\Sigma$
and $\mu$ are $3\times 3$ matrices, the latter two with dimensions of
mass. Finally, the VL leptons allows for additional Lagrangian terms,
given by
\begin{equation} \label{eq:VL}
  -\mathcal{L}_Y^{\text{VL}} =
  \sqrt{2} \, \overline{\Sigma} \, \lambda_L \, L_L \, \widetilde{H}^\dagger
  + \overline{e}_R \, \lambda_R \, L_L \, \widetilde{H}^\dagger
  + \overline{L}_L \, M_L \, L_R
  + \overline{\ell}_L  \, \epsilon \, L_R + \hc \, .
\end{equation}
Here $\lambda_L$ and $\lambda_R$ are dimensionless $3 \times 1$
vectors and $M_L$ a parameter with dimensions of mass. The $1 \times
3$ vector $\epsilon$ has dimensions of mass, and will be assumed to
vanish for simplicity.~\footnote{The $\epsilon$ term contributes to
the electron, muon and tau masses and is therefore constrained to be
small.}  The guiding principle when writing the Yukawa Lagragian in
Eq.~\eqref{eq:lag}, in particular the piece in Eq.~\eqref{eq:ISS3}, is
the conservation of lepton number, only allowed to be broken by the
$\overline{\Sigma^\prime}\,\mu\, \Sigma^{\prime c}$ term. In fact, in
the absence of the Majorana mass $\mu$, the Lagrangian would have an
additional $\rm U(1)_L$ global symmetry. In the following we will
consider $\mu \ll M_\Sigma$, corresponding to a \textit{slightly
  broken lepton number}, in the spirit of the original inverse seesaw
mechanism.~\footnote{In principle, a term of the form
$\overline{\Sigma}\,\mu^\prime\, \Sigma^c$ is also allowed by all
symmetries. However, it is well known that such a term would
contribute to neutrino masses in a subdominant way if $\mu$ and
$\mu^\prime$ are of the same order, see for
instance~\cite{CentellesChulia:2020dfh}. Therefore, we neglect this
term in the following.}

The scalar potential of the model is the same as in the SM,
\begin{equation} \label{eq:pot}
\mathcal V = m^2 |H|^2 + \lambda \, |H|^4 \, ,
\end{equation}
with $m^2$ a parameter with dimensions of [mass]$^2$. Therefore,
electroweak symmetry breaking takes place in the usual way, with
\begin{equation} \label{eq:EWvev}
\langle H \rangle = \frac{1}{\sqrt{2}} \, \left( \begin{array}{c}
0 \\
v
\end{array} \right) \, ,
\end{equation}
with $v \simeq 246$ GeV the SM Higgs vacuum expectation value. After
electroweak symmetry breaking, several terms in the Yukawa Lagrangian
in Eq.~\eqref{eq:lag} induce mixings in the neutral and charged lepton
sectors. In the bases $n \equiv n_L = \left(\nu_L,
(\Sigma^0)^c,(\Sigma^{\prime 0})^c,N_L,N_R^c\right)$,
$f_L=(e_L,(\Sigma^+)^c,(\Sigma^{\prime +})^c,E_L)$ and
$f_R=(e_R,\Sigma^-,\Sigma^{\prime -},E_R)$, the neutral and charged
fermion mass terms read
\begin{equation} \label{eq:massterms}
-\mathcal{L}_m =
\frac{1}{2} \overline{n^c} \, \mathcal M_N \, n
+ \overline{f_L} \, \mathcal M_C \,f_R + \, \hc \, ,
\end{equation}
with the mass matrices given by
\begin{equation} \label{eq:MN}
\mathcal M_N = \begin{pmatrix}
0 & m_D^T & 0 & 0 & 0 \\
m_D & 0 & M_\Sigma & m_L & 0 \\
0 & M_\Sigma^T & \mu & 0 & 0 \\
0 & m_L^T & 0 & 0 & M_L \\
0 & 0 & 0 & M_L & 0 \end{pmatrix} \, ,
\end{equation}
and
\begin{equation} \label{eq:MC}
\mathcal M_C = \begin{pmatrix}
m_e & \sqrt{2} \, m_D^T & 0 & 0 \\
0 & 0 & M_\Sigma & 0 \\
0 & M_\Sigma^T & \mu & 0 \\
m_R^T & \sqrt{2} \, m_L^T & 0 & - M_L \end{pmatrix} \, .
\end{equation}
Here we have defined
\begin{equation}
m_D = \frac{v}{\sqrt{2}} \, Y_\Sigma \, , \quad m_e = \frac{v}{\sqrt{2}} \, Y_e \, , \quad m_L = \frac{v}{\sqrt{2}} \, \lambda_L \quad \text{and} \quad m_R = \frac{v}{\sqrt{2}} \, \lambda_R \, .
\end{equation}
We note that the neutral lepton mass matrix $\mathcal M_N$ is $11
\times 11$, whereas the charged lepton mass matrix $\mathcal M_C$ is
$10 \times 10$.  They can be brought to diagonal form by means of the
unitary transformations $\mathcal U$, $\mathcal V^L$ and $\mathcal
V^R$, defined by
\begin{align}
  \mathcal{U}^* \, \mathcal{M}_N \, \mathcal{U}^\dagger &= \text{diag} \left( m_{N_i} \right) \, , \label{eq:diagN} \\
  \mathcal{V}^{L \, *} \, \mathcal{M}_C \, \mathcal{V}^{R \, \dagger} &= \text{diag} \left( m_{\chi_i} \right) \, , \label{eq:diagC}
\end{align}
resulting in the $11$ neutral (Majorana) fermion masses $m_{N_i}$ and
the $10$ charged (Dirac) fermion masses $m_{\chi_j}$, with
$i=1,\dots,11$ and $j=1,\dots,10$. In the following we will assume the
hierarchy of energy scales
\begin{equation} \label{eq:hier}
  \mu \ll m_D , m_L , m_R \ll M_\Sigma , M_L \, ,
\end{equation}
which allows one to obtain approximate expressions for the physical
lepton masses. We note that a small $\mu$ parameter is justified
through 't Hooft naturalness criterion~\cite{tHooft:1979rat}.  In the
case of the neutral leptons, one finds $3$ light states, to be
identified with the standard light neutrinos. Their mass matrix is
approximately given by
\begin{equation} \label{eq:mnu}
  m_\nu \approx m_D^T \, \left(M_\Sigma^T\right)^{-1} \mu \, M_\Sigma^{-1} \, m_D \, ,
\end{equation}
with corrections of the order of the small ratios
$\left(m_D/M_\Sigma\right)^2$ and $\left(m_L / M_L\right)^2$. This result is
proportional to the $\mu$ parameter. It is then clear that sizable
$Y_\Sigma$ Yukawa couplings and triplets at the TeV scale are
consistent with light neutrino masses due to the suppression by the
$\mu$ term. This is the inverse seesaw mechanism.

\section{Charged lepton anomalous magnetic moments}
\label{sec:g-2}

The charged lepton magnetic moments can be described by the effective
Hamiltonian~\cite{Crivellin:2018qmi}
\begin{equation} \label{eq:eff}
\mathcal{H} = c_{ij} \, \overline{\ell}_j \, \sigma_{\mu \nu} \, P_R \, \ell_i \, F^{\mu \nu} + \hc \, , 
\end{equation}
where $P_R = (1 + \gamma_5)/2$ is the usual right-handed chiral
projector, $F^{\mu \nu}$ the electromagnetic field strength tensor and
$\ell_i$ denote the light charged lepton mass eigenstates, equivalent
in our scenario to $\chi_{1,2,3}$. The anomalous magnetic moment is
given in terms of the real components of the diagonal $c$ coefficients
as
\begin{equation} \label{eq:relac}
  a_i = - \frac{2 \, m_i}{e} \left( c_{ii} + c_{ii}^* \right) = - \frac{4 \, m_i}{e} \, \text{Re} \, c_{ii} \, ,
\end{equation}
whereas the imaginary components would in turn induce electric dipole
moments.

\begin{figure}[tb]
\begin{subfigure}{0.45\linewidth}
  \centering
  \includegraphics[width=1.0\linewidth]{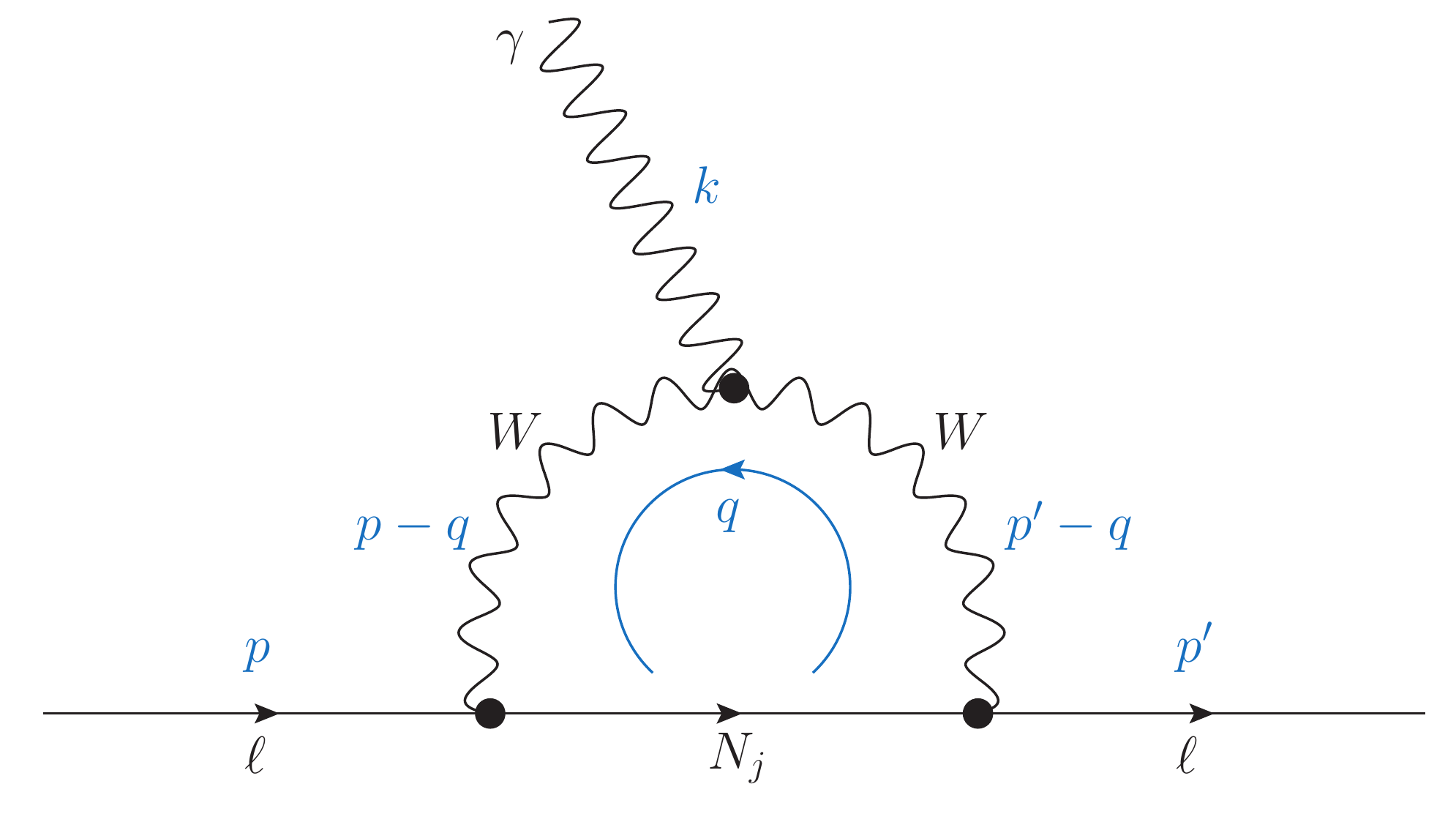}
  \caption{$W$ contribution}
  \label{fig:Wloop}
\end{subfigure}
\begin{subfigure}{0.45\linewidth}
  \centering
  \includegraphics[width=1.0\linewidth]{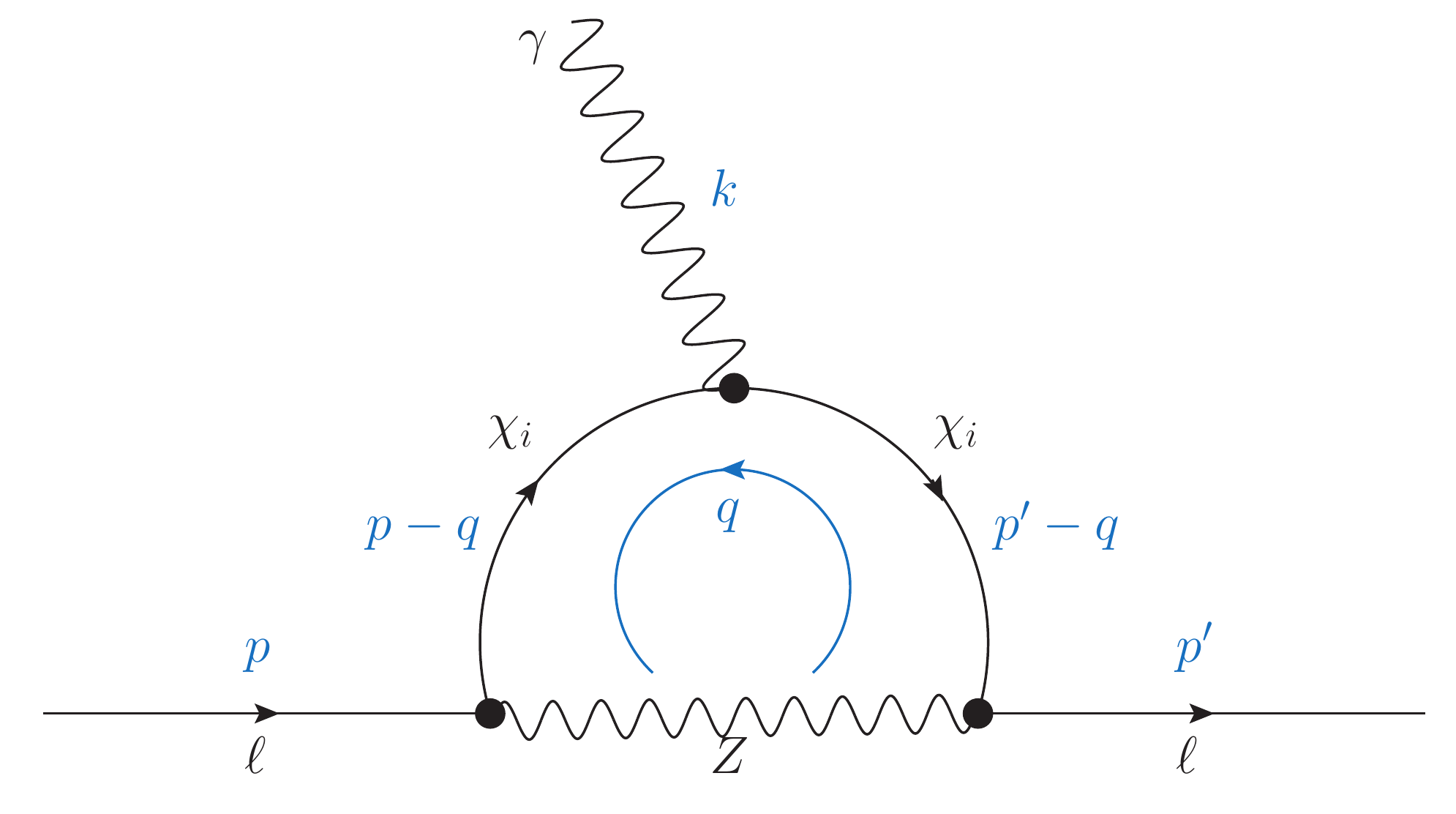}
  \caption{$Z$ contribution}
  \label{fig:Zloop}
\end{subfigure}
\newline
\centering
\begin{subfigure}{0.45\linewidth}
  \centering
  \includegraphics[width=1.0\linewidth]{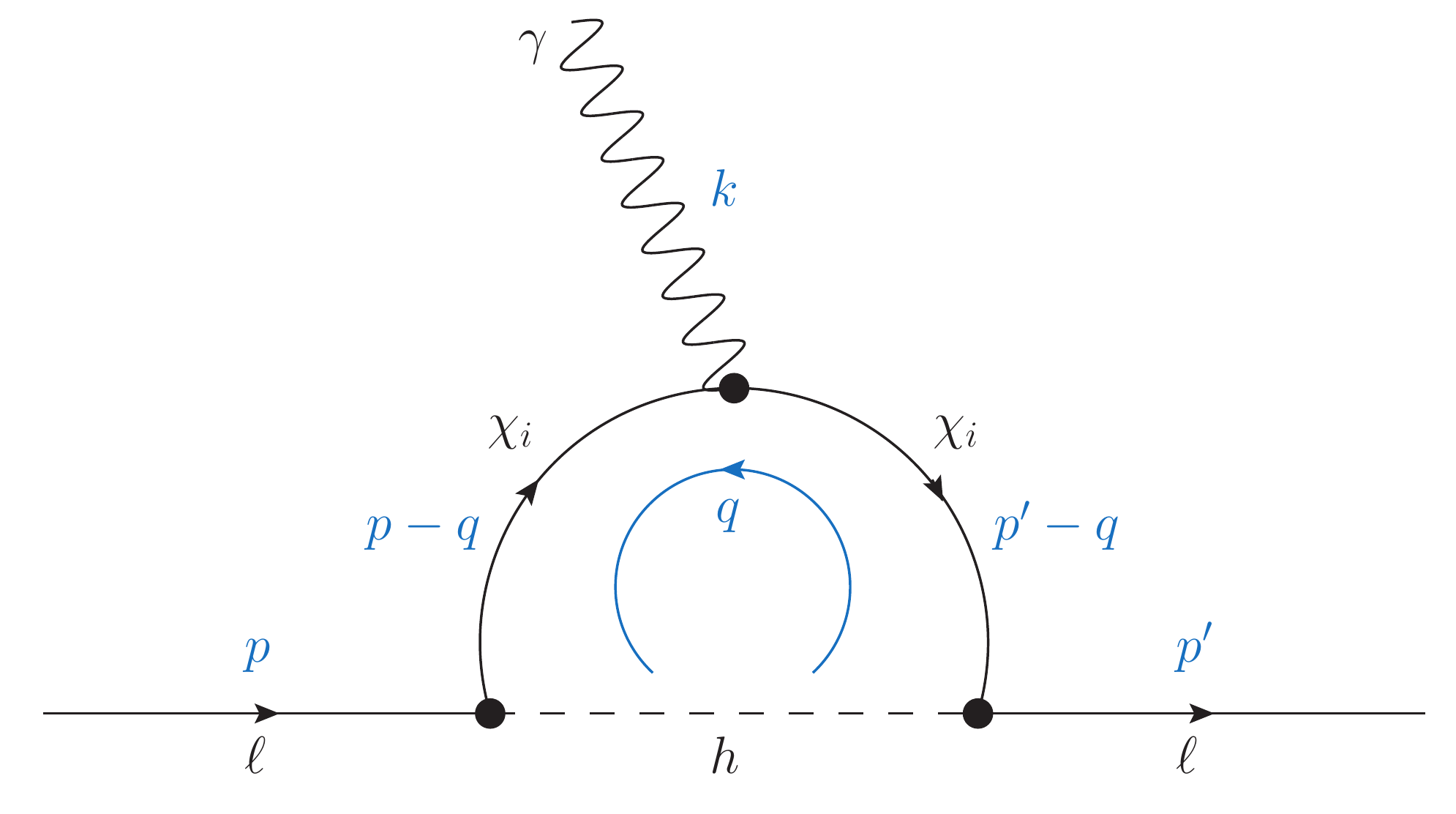}
  \caption{$h$ contribution}
  \label{fig:hloop}
\end{subfigure}
\caption{Feynman diagrams that contribute to the charged lepton
  anomalous magnetic moment at the 1-loop level in the ISS3VL. Here
  $\chi_i$ and $N_{j}$ denote any of the charged and neutral lepton
  mass eigenstates, respectively. Momenta are shown in blue.
  \label{fig:loops}
  }
\end{figure}

The ISS3VL has the ingredients to induce large charged lepton
anomalous magnetic moments, namely, light new particles with sizable
couplings to the charged leptons. In the ISS3VL, new contributions to
the charged lepton anomalous magnetic moments are induced at the
1-loop level, as shown in Fig.~\ref{fig:loops}. While these diagrams
also exist in the SM, in the ISS3VL the mass eigenstates $N_i$ and
$\chi_i$ include new heavy states beyond the SM leptons. Moreover, the
couplings of the SM states get modified due to mixings with the new
BSM states. The amplitudes of these Feynman diagrams are given
by~\footnote{In order to obtain the correct sign for the ISS3VL
  contributions to the electron and muon $g-2$ one must use a
  consistent set of sign conventions for the Feynman rules of the
  model. We used the useful Ref.~\cite{Romao:2012pq} to guarantee the
  consistency of the amplitudes in
  Eqs.~\eqref{eq:ampW}-\eqref{eq:amph}.}
\begin{align}
  i \, \mathcal{M}_\ell^W =& \int\dfrac{d^4q}{\left(2\pi\right)^4}\bar{u}_{\ell }\left(p^\prime\right) \, i \, \gamma^\beta \left[\left(g^L_{\chi N W}\right)_{\ell j}P_L+\left(g^R_{\chi N W}\right)_{\ell j}P_R\right]i\dfrac{\slashed{q}+m_{N_j}}{q^2-m^2_{N_j}} \nonumber \\
    & i \, \gamma^\alpha \left[\left(g^{L}_{\chi N W}\right)_{j\ell} P_L + \left(g^R_{\chi N W}\right)_{j\ell} P_R\right] \dfrac{-i\left(g_{\alpha\tilde{\alpha}}-\frac{\left(p-q\right)_\alpha \left(p-q\right)_{\tilde{\alpha}}}{m_W^2}\right)}{\left(p-q\right)^2-m_W^2} \, i \, \Gamma^{\mu\tilde{\alpha}\tilde{\beta}} \nonumber \\ 
  & \dfrac{-i\left(g_{\beta\tilde{\beta}}-\frac{\left(p^\prime-q\right)_\beta\left(p^\prime- q\right)_{\tilde{\beta}}}{m_W^2}\right)}{\left(p^\prime-q\right)^2-m_W^2}u_\ell \left( p \right) \varepsilon_\mu \, , \label{eq:ampW}
\end{align}
\begin{align}
  i \, \mathcal{M}_\ell^Z =& \int\dfrac{d^4q}{\left(2\pi\right)^4}\bar{u}_{\ell }\left(p^\prime\right) \, i \, \gamma^\beta\left[\left(g^L_{\chi Z}\right)_{i\ell }P_L+\left(g^R_{\chi Z}\right)_{i\ell }P_R\right] i\dfrac{\left(\slashed{p}^\prime-\slashed{q}\right)+m_{\chi_i}}{\left(p^\prime-q\right)^2-m^2_{\chi_i}} \nonumber \\
  & i \, \gamma^\mu \left[\left(g^L_{\chi \gamma}\right)_{ii}P_L+\left(g^R_{\chi \gamma}\right)_{ii}P_R\right] i\dfrac{\left(\slashed{p}-\slashed{q}\right)+m_{\chi_i}}{\left(p-q\right)^2-m^2_{\chi_i}}  \, i \, \gamma^\alpha \left[\left(g^L_{\chi Z}\right)_{\ell i}P_L+\left(g^R_{\chi Z}\right)_{2i}P_R\right] \nonumber \\
  & \dfrac{-i\left(g_{\alpha\beta}-\frac{q_\alpha q_\beta}{m_Z^2}\right)}{q^2-m_Z^2}u_\ell \left( p \right) \varepsilon_\mu \, , \label{eq:ampZ} 
  \\
  i \, \mathcal{M}_\ell^h =& \int \dfrac{d^4 q}{\left( 2 \pi \right)^4} \bar{u}_{\ell} \left( p^\prime \right) \, i \, \left[ \left( g^L_{\chi h} \right)_{i \ell} P_L + \left( g^R_{\chi h} \right)_{i \ell} P_R \right] i \dfrac{\left( \slashed{p}^\prime - \slashed{q} \right) + m_{\chi_i}}{\left( p^\prime - q \right)^2 - m^2_{\chi_i}} \nonumber \\
  & i \, \gamma^\mu \left[\left(g^L_{\chi \gamma}\right)_{ii}P_L+\left(g^R_{\chi \gamma}\right)_{ii}P_R\right] i \dfrac{\left( \slashed{p} - \slashed{q} \right) + m_{\chi_i}}{\left( p - q \right)^2 - m^2_{\chi_i}} \, i \, \left[ \left( g^L_{\chi h}\right)_{\ell i} P_L + \left( g^R_{\chi h}\right)_{\ell i} P_R \right] \nonumber \\
  & \dfrac{i}{q^2 - m_h^2} u_\ell \left( p \right) \varepsilon_\mu \, . \label{eq:amph}
  \end{align}
Here $P_L = (1 - \gamma_5)/2$ is the left-handed chiral projector,
$\varepsilon_\mu$ is the photon polarization 4-vector and the
couplings $g^{L,R}_{\chi N W}$, $g^{L,R}_{\chi Z}$, $g^{L,R}_{\chi
  \gamma}$, $g^{L,R}_{\chi h}$ and $\Gamma$ are defined in
Appendix~\ref{sec:app1}. A sum over the indices $i,j$ is implicit in
these expressions, while $\ell$ is the index of the external charged
lepton. We have computed the amplitudes in
Eqs.~\eqref{eq:ampW}-\eqref{eq:amph} with the help of {\tt
  Package-X}~\cite{Patel:2015tea}. After projecting onto the operator
in Eq.~\eqref{eq:eff}, one obtains analytical expressions for the
contributions to the $c$ coefficient, which can then be translated
into contributions to the charged leptons $g-2$ thanks to the relation
in Eq.~\eqref{eq:relac}. The total ISS3VL contribution to $\Delta
a_{\ell}$ can be written as~\footnote{Higher-order contributions, such
  as those induced by 2-loop Barr-Zee diagrams~\cite{Barr:1990vd},
  will be neglected in the following.}
\begin{equation}
\Delta a_{\ell}  = \Delta a_{\ell} \left( W \right) + \Delta a_{\ell} \left( Z \right) + \Delta a_{\ell} \left( h \right) \, .
\end{equation}
Assuming that the ISS3VL states $N_i$ and $\chi_i$ are much heavier
than the SM states (this is, assuming the mass hierarchy $m_{N_i},
m_{\chi_i} \gg m_h, m_W, m_Z \gg m_\ell$), one can find approximate
expressions for the three contributions:
\begin{align}
  \Delta a_{\ell} \left( W \right) \simeq & \, \dfrac{m_\ell}{32 \, \pi^2 \, m_W^2} \left\{ \frac{4}{3} \, m_\ell \left[ 1 - \frac{3 \, m_W^2}{4 \, m_{N_i}^2} \left(11 + 6 \log \frac{m_W^2}{m_{N_i}^2} \right) \right] \left( C^2_{\chi N W} \right)_{i \ell} \right. \nonumber \\ 
  & \left. - \, m_{N_i} \left[ 1 - \frac{3 \, m_W^2}{m_{N_i}^2} \left( 3 + 2 \log \frac{m_W^2}{m_{N_i}^2} \right) \right] \left( D^2_{\chi N W} \right)_{i \ell} \right\} \, , \label{eq:DeltaW} \\
  \Delta a_{\ell} \left( Z \right) \simeq & \, \dfrac{m_\ell}{32 \, \pi^2 \, m_Z^2} \Biggl\{ - \frac{5}{3} \, m_\ell \left( C_{\chi Z}^2 \right)_{i \ell} + \, m_{\chi_i} \left( D_{\chi Z}^2 \right)_{i \ell} \Biggr\} \, , \label{eq:DeltaZ} \\
  \Delta a_{\ell} \left( h \right) \simeq & \, \dfrac{m_\ell}{32 \, \pi^2 \, m_{\chi_i}^2} \Biggl\{ \frac{1}{3} \, m_\ell \left( C_{\chi h}^2 \right)_{i \ell} + m_{\chi_i} \left( D_{\chi h}^2 \right)_{i \ell} \Biggr\} \, . \label{eq:Deltah}
\end{align}
Here we have defined the coupling combinations
\begin{equation}
  \left( C_Y^2 \right)_{i \ell} \equiv \left| \left( g^L_Y \right)_{i \ell} \right|^2 + \left| \left( g^R_Y \right)_{i \ell} \right|^2 \qquad \mathrm{and} \qquad \left( D_Y^2 \right)_{i \ell} \equiv \left( g^L_Y \right)_{i \ell} \, \left( g^R_Y \right)_{i \ell}^* + \left( g^L_Y \right)_{i \ell}^* \, \left( g^R_Y \right)_{i \ell} \, ,
\end{equation}
with $Y = \chi N W, \chi Z, \chi h$. Again, the indices $i$ and $\ell$
denote the BSM particle running in the loop and the charged lepton,
respectively. We have checked that Eqs.~\eqref{eq:DeltaW},
\eqref{eq:DeltaZ} and \eqref{eq:Deltah} reproduce the ISS3VL
contributions to the charged leptons anomalous magnetic moments in
very good approximation. Nevertheless, full expressions are given in
Appendix~\ref{sec:app2} and used in the numerical analysis presented
in the next Section. We note that $\Delta a_\ell(W)$, $\Delta
a_\ell(Z)$ and $\Delta a_\ell(h)$ contain contributions proportional
to $m_N \, g_{\chi N W}^L \, g_{\chi N W}^R$, $m_\chi \, g_{\chi Z}^L
\, g_{\chi Z}^R$ and $m_\chi \, g_{\chi h}^L \, g_{\chi h}^R$,
respectively, proportional to the mass of the fermion in the
loop. These terms are usually called \textit{chirally-enhanced
  contributions} and they typically dominate due to the large masses
of the heavy fermions running in the loop.

\section{Phenomenological discussion}
\label{sec:pheno}

We proceed now to present our phenomenological exploration of the
parameter space of the ISS3VL.

\subsection{Experimental constraints}
\label{subsec:bounds}

Let us first discuss how we fix the parameters of the model in order
to reproduce the measured lepton masses and mixings. The $Y_e$ Yukawa
matrix will be fixed to the same values as in the SM, hence neglecting
corrections from the mixing between the SM charged lepton states and
the charged components of the $\Sigma$ and $\Sigma^\prime$
triplets. These corrections are multiplicative and enter at order
$\sim \left( m_D / M_\Sigma \right)^2$ and can thus be safely
neglected. The same argument applies to the mixing with the charged
components of the VL leptons, which enter at order $\sim \left( m_R /
M_L \right)^2$. Without loss of generality, we will work in the basis
in which $M_\Sigma$ is diagonal and $\mu$ is a general complex
symmetric matrix. In this case, $Y_\Sigma$ and $\mu$ must be properly
fixed in order to reproduce neutrino oscillation
data~\cite{deSalas:2020pgw}. In principle, one can fix the entries of
$\mu$ to some input values and express the $Y_\Sigma$ Yukawa matrix by
means of the master
parametrization~\cite{Cordero-Carrion:2018xre,Cordero-Carrion:2019qtu},
which in this case reduces to a modified Casas-Ibarra
parametrization~\cite{Casas:2001sr}. While this is perfectly valid,
one generically obtains $Y_\Sigma$ matrices with sizable off-diagonal
entries unless some input parameters are tuned very finely. Due to the
strong constraints from the non-observation of lepton flavor violating
processes, this excludes most of the parameter points. Therefore, we
take the alternative choice of fixing $Y_\Sigma$ to specific input
values, diagonal for simplicity, and computing $\mu$ by inverting
Eq.~\eqref{eq:mnu} as
\begin{equation} \label{eq:param}
  \mu = M_\Sigma^T \, \left(m_D^T\right)^{-1} m_\nu \, m_D^{-1} \, M_\Sigma \, ,
\end{equation}
where $m_\nu = U_\nu^* \, \widehat m_\nu \, U_\nu^\dagger$. Here
$U_\nu$ is the leptonic mixing matrix measured in oscillation
experiments, given in terms of $3$ mixing angles and $3$ CP-violating
phases, while $\widehat m_\nu$ is a diagonal matrix containing the
physical neutrino mass eigenvalues. Eq.~\eqref{eq:param} guarantees
that all the parameter points considered in our numerical analysis are
compatible with neutrino oscillation data. In our analysis we use the
results of the global fit in~\cite{deSalas:2020pgw} and we consider
both normal and inverted neutrino mass orderings.

In order to ensure compatibility with constraints from flavor and
electroweak precision data we use the bounds derived
in~\cite{Biggio:2019eeo}, where a global analysis is performed in the
context of general type-III seesaw models. The limits provided in this
reference are given for the $3 \times 3$ matrix $\eta$, defined in our
case in terms of the matrices
\begin{equation}
  M_D = \left( \begin{array}{c}
                 m_D \\
                 0
               \end{array}\right)  \quad \text{and} \quad 
  M = \left( \begin{array}{cc}
                 0 & M_\Sigma \\
                 M_\Sigma^T & \mu
               \end{array}\right) \, ,
\end{equation}
as
\begin{align}
  \eta &= \frac{1}{2} M_D^\dagger \, \left(M^\dagger \right)^{-1} M^{-1} \, M_D \nonumber \\
  &= \frac{1}{2} m_D^\dagger \, \left(M_\Sigma^\dagger \right)^{-1} \left[ \mathbb{I}_3 + \mu^* \left(M_\Sigma^* \right)^{-1} \left(M_\Sigma^T \right)^{-1} \mu \right] M_\Sigma^{-1} \, m_D \\
  &\approx \frac{1}{2} m_D^\dagger \, \left(M_\Sigma^\dagger \right)^{-1} M_\Sigma^{-1} \, m_D \, . \nonumber
\end{align}
In our analysis, we make sure that the bounds are respected by
computing the $\eta$ matrix in all the parameter points considered. As
we will explain below, these limits imply very small BSM contributions
in the ISS3, thus motivating our ISS3VL extension. Furthermore, we
also consider the decay widths for the processes $Z \to \ell^+ \ell^-$
and $h \to \ell^+ \ell^-$, with $\ell = e, \mu$, which are also
affected due to the mixing of the light charged leptons with the heavy
states in our model. They are computed as
\begin{align}
  \Gamma(Z \to \ell^+ \ell^-) &= \frac{m_Z^3}{12 \pi v^2} \left[ \left| \left(g^V_{\chi Z}\right)_{\ell \ell} \right|^2 + \left| \left(g^A_{\chi Z}\right)_{\ell \ell} \right|^2 \right] \, , \\
  \Gamma(h \to \ell^+ \ell^-) &= \frac{m_h}{8 \pi} \left[ \left| \left(g^L_{\chi h}\right)_{\ell \ell} \right|^2 + \left| \left(g^R_{\chi h}\right)_{\ell \ell} \right|^2 \right] \, ,
\end{align}
with $g^V_{\chi Z} = g^L_{\chi Z} + g^R_{\chi Z}$ and $g^A_{\chi Z} =
g^R_{\chi Z} - g^L_{\chi Z}$. The $Z \to \ell^+ \ell^-$ decay turns
out to provide an important constraint in our setup. In fact, it has
been recently pointed out that this process potentially correlates
with the charged leptons $g-2$~\cite{Crivellin:2021rbq}. We define the
ratios
\begin{equation}
  R_{Z \ell \ell} = \frac{\Gamma(Z \to \ell^+ \ell^-)}{\Gamma_{\rm SM}(Z \to \ell^+ \ell^-)} \, ,
\end{equation}
with $\Gamma_{\rm SM}(Z \to \ell^+ \ell^-)$ the SM predicted decay
width, and impose that $R_{Z \ell \ell}$ lies within the 95\% CL
range, which we estimate to be $0.995 < R_{Z e e} < 1.003$ and $0.993
< R_{Z \mu \mu} < 1.006$~\cite{Zyla:2020zbs}. Regarding the Higgs
boson decays, no constraints are actually obtained from them, since at
present there is no hint for $h \to e^+ e^-$ and evidence for $h \to
\mu^+ \mu^-$ was only obtained
recently~\cite{Sirunyan:2020two}. Therefore, they will be considered
as predicted observables, potentially correlated with $\Delta
a_\ell$~\cite{Fajfer:2021cxa,Crivellin:2021rbq}.

Finally, we also impose bounds from collider searches. The type-III
seesaw triplets have been searched for at the LHC in multilepton final
states, both by ATLAS~\cite{Aad:2020fzq} and
CMS~\cite{CMS:2012ra,Sirunyan:2017qkz}. No excess above the expected
SM backgrounds has been found, hence allowing the experimental
collaborations to set limits on the triplet mass and couplings. Using
a data sample obtained with proton collisions at $\sqrt{s} = 13$ TeV
and an integrated luminosity of $35.9$ fb$^{-1}$, CMS reports a lower
bound on the triplet mass of $840$ GeV at $95 \%$ confidence level, if
the triplet couplings are assumed to be lepton flavor
universal~\cite{Sirunyan:2017qkz}. While the flavor structure of the
triplet couplings to leptons does not affect the heavy triplet pair
production cross-sections, driven by gauge interactions, it has an
impact on the flavor composition of the multilepton signature. The
limit changes if the assumption of lepton flavor universal couplings
is dropped, resulting in a more stringent bound when the triplet
couples mainly to electrons and a more relaxed bound when the triplet
couples mainly to taus, with values ranging between $390$ and $930$
GeV. The bounds from the CMS collaboration in~\cite{Sirunyan:2017qkz}
are applied in our analysis. However, since the CMS analysis focuses
on the standard type-III seesaw scenario, and does not consider the
particular features of the ISS3VL model, several simplifying
assumptions must be made. We define
\begin{equation}
  B_{A \alpha} = \frac{\Gamma(\Sigma_A^0 \to \ell_\alpha + \text{boson}) + \Gamma(\Sigma_A^+ \to \ell_\alpha + \text{boson})}{\sum_\alpha \left[ \Gamma(\Sigma_A^0 \to \ell_\alpha + \text{boson}) + \Gamma(\Sigma_A^+ \to \ell_\alpha + \text{boson}) \right]} \, ,
\end{equation}
where $\Sigma^0_A$ and $\Sigma_A^+$ are the quasi-Dirac pairs
approximately formed by the mass eigenstates $N_i + N_j$ and $\chi_i +
\chi_j$, respectively.  An implicit sum over the bosons in the final
states is also assumed, including decays to $W^\pm$, $Z$ and $h$. For
instance, in a parameter point in which the lightest BSM states are
mainly composed by the components of the $\Sigma_1$ triplet, we have
$i=4$, $j=5$ and $\Gamma(\Sigma^0_1 \to \ell_\alpha + \text{boson})
\equiv \Gamma(N_4 \to W^\pm \ell^\mp) + \Gamma(N_4 \to Z \nu_\alpha) +
\Gamma(N_4 \to h \nu_\alpha) + \Gamma(N_5 \to W^\pm \ell^\mp) +
\Gamma(N_5 \to Z \nu_\alpha) + \Gamma(N_5 \to h \nu_\alpha)$. We note
that $B_{Ae} + B_{A\mu} + B_{A\tau} = 1$. This is the quantity that we
use to confront each quasi-Dirac pair with the limits given on Figure
3 of~\cite{Sirunyan:2017qkz}. Our approach approximates the total
heavy triplet pair production to $pp \to \Sigma_A^0 \Sigma_A^+$, which
is known to give the dominant contribution at the
LHC~\cite{Biggio:2011ja}. Furthermore, we apply two additional
simplifications. First, since CMS assumes the neutral and charged
components of the triplet to be mass degenerate, we adopt a
conservative approach and take the lowest of them as the triplet mass
to be used in the analysis. And second, we do not apply the CMS bounds
to quasi-Dirac triplet pairs that are largely mixed with the VL
leptons, since their production cross-section is clearly reduced with
respect to the pure triplet case.~\footnote{In practice, we do not
  apply the CMS bounds in cases with large mixings. For instance, they
  are not applied to $\chi_i$ Dirac states that combine a left-handed
  fermion that is mostly a type-III triplet with a right-handed
  fermion that is mostly a VL lepton, or vice versa.} We believe that
our assumptions conservatively adapt the CMS limits
in~\cite{Sirunyan:2017qkz} to our scenario. We note that ATLAS finds a
similar bound on the triplet mass in the flavor universal scenario,
ruling out (at $95 \%$ confidence level) values below $790$
GeV~\cite{Aad:2020fzq}. Finally, LHC limits on VL leptons strongly
depend on their decay modes, namely the flavor of the charged leptons
produced in the final states~\cite{Falkowski:2013jya}. In our analysis
we will consider $M_L \geq 500$ GeV, a conservative value that
guarantees compatibility with current LHC searches. These limits are
expected to be improved by the end of the LHC
Run-III~\cite{Freitas:2020ttd}.

\subsection[\texorpdfstring{$(g-2)_{e,\mu}$ in the ISS3}{g-2 in the ISS3}]{$\boldsymbol{(g-2)_{e,\mu}}$ in the ISS3}
\label{subsec:ISS3pheno}

\begin{figure}[!tb]
  \centering
  \includegraphics[width=0.7\linewidth]{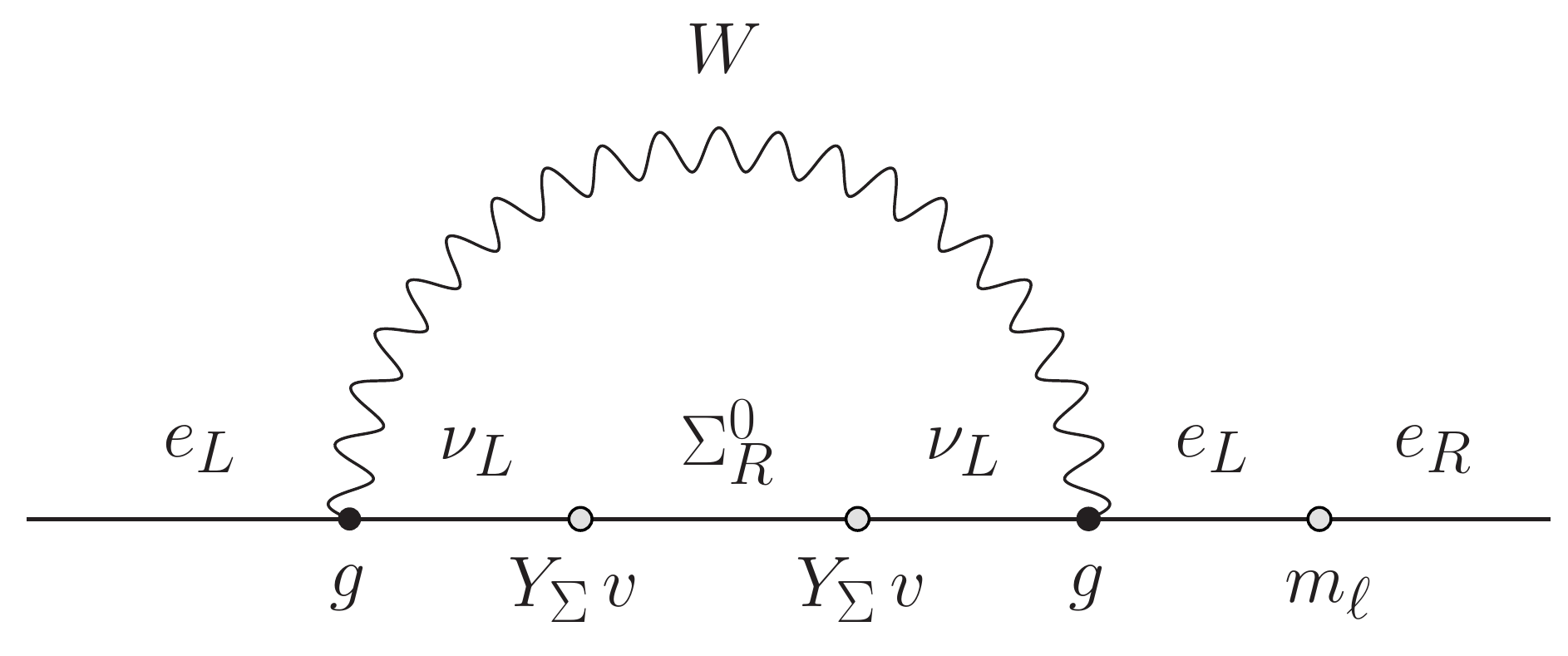}
  \caption{Dominant $W$ contribution in the ISS3. Mass insertions are represented by white blobs.
    \label{fig:WISS3}
    }
\end{figure}

Before studying the electron and muon $g-2$ in the ISS3VL, let us
discuss these observables in the context of the \textit{pure} ISS3 and
show that this model is unable to address the existing
discrepancies. One can easily reach this conclusion by estimating the
size of the dominant contributions to the charged lepton
$g-2$. Fig.~\ref{fig:WISS3} shows the dominant $W$
contribution. Assuming that the chirally-enhanced term in
Eq.~\eqref{eq:DeltaW} dominates, one finds the estimate
\begin{equation} \label{eq:estimISS3}
  \Delta a_\ell(W) \sim - \frac{1}{32 \pi^2} \frac{m_N \, m_\ell}{m_W^2} \, g^2 \, \left( \frac{m_D}{M_\Sigma} \right)_{\ell\ell}^2 \, \frac{m_\ell}{v} \sim - 10^{-3} \, \frac{m_N \, m_\ell^2}{m_W^3} \, \eta_{\ell \ell} \, .
\end{equation}
First of all, we notice that this contribution is always negative
since $\eta_{\ell \ell} > 0$. Therefore, it cannot accommodate the
muon $g-2$ anomaly, which requires $\Delta a_\mu > 0$. This result was
already found in early studies of the charged leptons anomalous
magnetic moments in seesaw
scenarios~\cite{Biggio:2008in,Chao:2008iw,Chua:2010me}, as well as
in~\cite{Freitas:2014pua}. Furthermore, the absolute value of $\Delta
a_\ell(W)$ is also too small to account for the anomalies. This
implies that the electron $g-2$ cannot be explained either in the
ISS3. In this regard, we highlight the relevance of the $m_\ell/v$
factor in Eq.~\eqref{eq:estimISS3}. This factor is not apparent when
inspecting the analytical expressions for the couplings in
Appendix~\ref{sec:app1}. In fact, the individual contributions to
$\Delta a_\ell(W)$ by the neutral fermions in the loop are larger than
their sum, $\Delta a_\ell(W)$, by a factor $\sim v/m_\ell$. Therefore,
a strong cancellation among them takes place. This cancellation can be
easily understood due to the chirality-flipping nature of the dipole
moment operator in Eq.~\eqref{eq:eff}. The factor $m_\ell/v$ is
required to flip the chirality of the fermion line and induce a
contribution to a dipole moment. One can now consider $m_N = 1$ TeV to
obtain
\begin{align}
  \Delta a_e(W) &\sim -5 \cdot 10^{-13} \, \eta_{ee} \, , \\
  \Delta a_\mu(W) &\sim -2 \cdot 10^{-8} \, \eta_{\mu\mu} \, .
\end{align}
Since $\eta_{ee}$ and $\eta_{\mu \mu}$ are constrained to be smaller
than $\sim 10^{-4}$~\cite{Biggio:2019eeo}, these contributions fail to
address the electron and muon $g-2$ anomalies by several orders of
magnitude. The same argument can be applied to the $Z$ and $h$
contributions to find that they are actually even more suppressed. In
summary, the suppression by the $m_\ell/m_W$ chirality flip and the
stringent bounds on $\eta_{\ell \ell}$ imply that the ISS3 cannot
induce sizable contributions. This, added to the fact that the
contributions to the muon $g-2$ have the wrong sign, implies that the
ISS3 cannot explain the deviations in the electron and muon anomalous
magnetic moments. We now proceed to show that the additional
ingredients in our extended model can alter this conclusion.

\subsection[\texorpdfstring{$(g-2)_{e,\mu}$ in the ISS3VL}{g-2 in the ISS3VL}]{$\boldsymbol{(g-2)_{e,\mu}}$ in the ISS3VL}
\label{subsec:ISS3VLpheno}

\begin{figure}[!tb]
  \centering
  \includegraphics[width=0.7\linewidth]{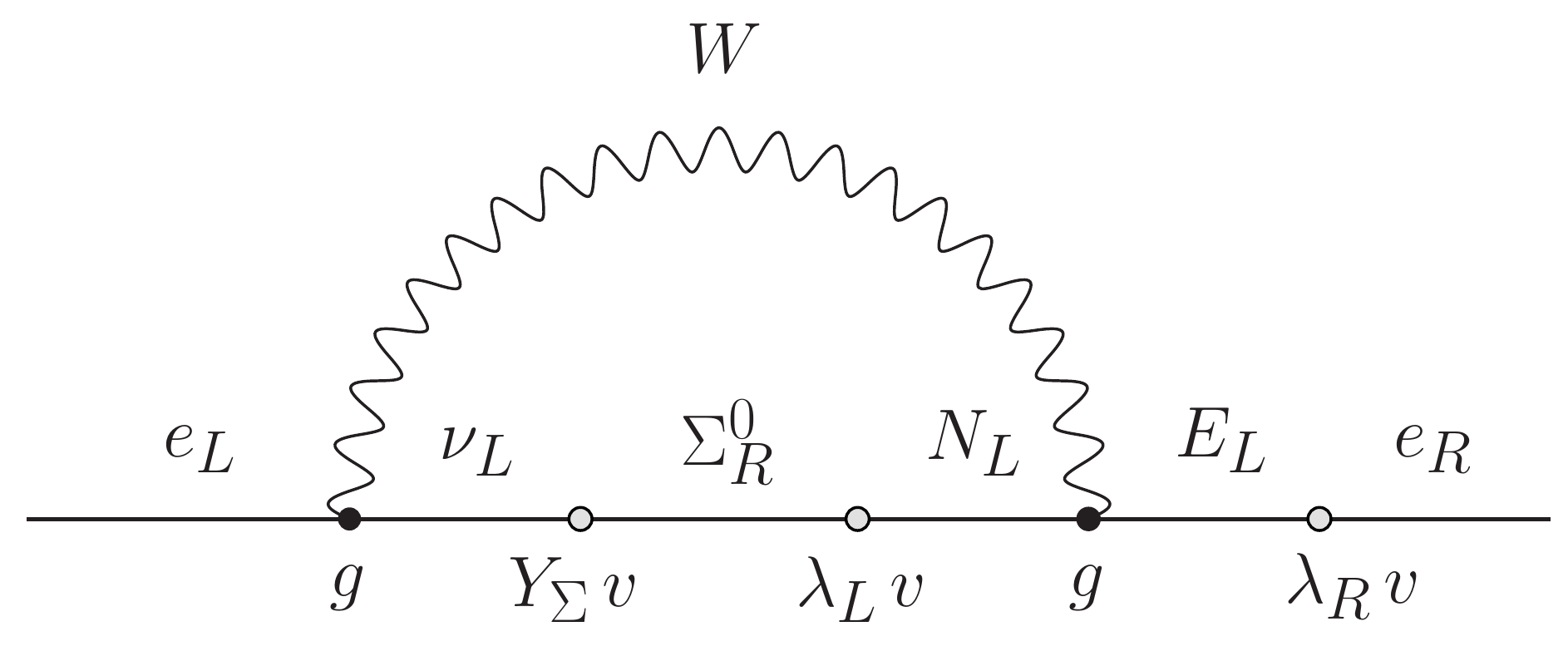}
  \caption{Dominant $W$ contribution in the ISS3VL. Mass insertions are represented by white blobs.
    \label{fig:WISS3VL}
    }
\end{figure}

As already discussed, the ISS3 cannot explain the experimental
anomalies in the electron and muon anomalous magnetic
moments. Therefore, we now consider its ISS3VL extension. In this case
one has $W$ contributions such as the one shown in
Fig.~\ref{fig:WISS3VL}. We can now derive an analogous estimate, along
the same lines as in the case of the ISS3. One finds
\begin{equation} \label{eq:estimISS3VL}
  |\Delta a_\ell(W)| \sim \frac{1}{32 \pi^2} \frac{m_N \, m_\ell}{m_W^2} \, g^2 \, \left( \frac{m_D}{M_\Sigma} \right)_{\ell\ell} \frac{m_L}{M_L} \, \, \frac{m_R}{M_L} \sim 10^{-3} \, \frac{m_N \, m_\ell \, m_L \, m_R}{m_W^2 \, M_L^2} \, \sqrt{\eta_{\ell \ell}} \, .
\end{equation}
One can now choose $m_N = 1$ TeV, $M_L = 500$ GeV, $m_L = 200$ GeV and
$m_R = 10$ GeV to obtain
\begin{align}
  |\Delta a_e(W)| &\sim 6 \cdot 10^{-10} \, \sqrt{\eta_{ee}} \, , \\
  |\Delta a_\mu(W)| &\sim 10^{-7} \, \sqrt{\eta_{\mu\mu}} \, .
\end{align}
Therefore, even after the suppression given by $\sqrt{\eta_{\ell\ell}}
\lesssim 10^{-2}$ these $W$ contributions can address the current
discrepancies with the electron and muon $g-2$ measurements.
Furthermore, the signs of these contributions are not fixed and can be
properly adjusted by fixing the signs of the relevant Yukawa
couplings. We note that the loop in Fig.~\ref{fig:WISS3VL} is
proportional to the product $Y_\Sigma \, \lambda_L \, \lambda_R$
which, as shown below, will be crucial for the resulting values for
$\Delta a_\ell$ in the ISS3VL model. Similar $h$ contributions are
also found, again proportional to the $Y_\Sigma \, \lambda_L \,
\lambda_R$ product. Therefore, the model is in principle capable of
producing sizable contributions to the electron and muon $g-2$. We now
proceed to confirm this by performing a detailed numerical analysis of
the parameter space of the model. Since we are interested in $\Delta
a_e$ and $\Delta a_\mu$, we fix $\left(\lambda_L\right)_3 =
\left(\lambda_R\right)_3 = 0$ and randomly scan within the following
parameter ranges:
\begin{center}
  {
    \renewcommand{\arraystretch}{1.3}
    \begin{tabular}{ccc}
      \hline
      {\bf Parameter} & {\bf Min} & {\bf Max} \\
      \hline
      $\left(M_\Sigma\right)_{ii}$ & $850$ GeV & $1.5$ TeV \\
      $M_L$ & $500$ GeV & $1.5$ TeV \\
      $\left(Y_\Sigma\right)_{ii}$ & $0.05$ & $0.2$ \\
      $\left(\lambda_L\right)_1$ & $-\sqrt{4 \pi}$ & $-0.1$ \\
      $\left(\lambda_L\right)_2$ & $0.1$ & $\sqrt{4 \pi}$ \\
      $\left(\lambda_R\right)_1$ & $0.05$ & $0.5$ \\
      $\left(\lambda_R\right)_2$ & $0.05$ & $0.5$ \\
      \hline
    \end{tabular}
  }
\end{center}
Some comments about the chosen ranges are in order. First, the ranges
for the mass parameters $\left(M_\Sigma\right)_{ii}$ and $M_L$ have
been selected following the discussion on LHC bounds of
Sec.~\ref{subsec:bounds}. Many of the parameter points in our scan
were ruled out due to LHC searches for triplets, but we also find that
a substantial fraction pass the test. The ranges for the Yukawa
couplings have been chosen in order to maximize the resulting $\Delta
a_\ell$. The usual ISS3 Yukawas $\left(Y_\Sigma\right)_{ii}$ have been
scanned around their maximal values compatible with the $\eta_{ii}$
bounds. A relative sign between $\left(\lambda_L\right)_1$ and
$\left(\lambda_L\right)_2$ has been introduced in order to obtain
$\Delta a_e < 0$ and $\Delta a_\mu > 0$, as required by the
experimental hints. Finally, the corrections to $\Gamma(Z \to \ell^+
\ell^-)$ tend to be too large unless $\left(\lambda_R\right)_{1,2}
\lesssim 0.5$.

Our results are based on a random scan with $50.000$ parameter points,
out of which about $12\%-13\%$ pass all the experimental tests. We
have selected normal neutrino mass ordering. However, we have also run
a second scan with inverted ordering and found the same qualitative
results. As already explained, we consider a scenario with diagonal
$Y_\Sigma$ and $M_\Sigma$ matrices. In this case, the lepton mixing
angles encoded in the matrix $U_\nu$ are generated by the off-diagonal
entries of the $\mu$ matrix and all lepton flavor violating processes
are strongly suppressed. For this reason, the bounds on the
$\eta_{ij}$ entries, with $i \neq j$, are easily satisfied. In
constrast, the bounds on the diagonal elements of the $\eta$ matrix
turn out to be very important, removing a significant amount of the
parameter points considered and implying the approximate bounds
$\left(Y_\Sigma\right)_{11} \lesssim 0.2$ and
$\left(Y_\Sigma\right)_{22} \lesssim 0.15$ for triplet masses of the
order of the TeV. Another very important constraint in our setup is
provided by the decay $Z \to \ell^+ \ell^-$. The mixing between the SM
charged leptons and the new charged BSM states from the $\Sigma$ and
$\Sigma^\prime$ triplets and $L_{L,R}$ VL doublets reduces $\Gamma(Z
\to \ell^+ \ell^-)$ with respect to its SM value. This has a strong
impact on the $m_D/M_\Sigma$ and $m_R/M_L$ ratios. Since these ratios
must be sizable in order to induce large contributions to $\Delta
a_\ell$, see Fig.~\ref{fig:WISS3VL}, this limit is crucial for the
correct evaluation of our scenario. Finally, the CMS limits discussed
in Sec.~\ref{subsec:bounds} also have an impact, discarding some of
the parameter points in our scan.

\begin{figure}[!tb]
  \centering
  \begin{subfigure}{0.48\textwidth}
    \centering  
    \includegraphics[width=\linewidth]{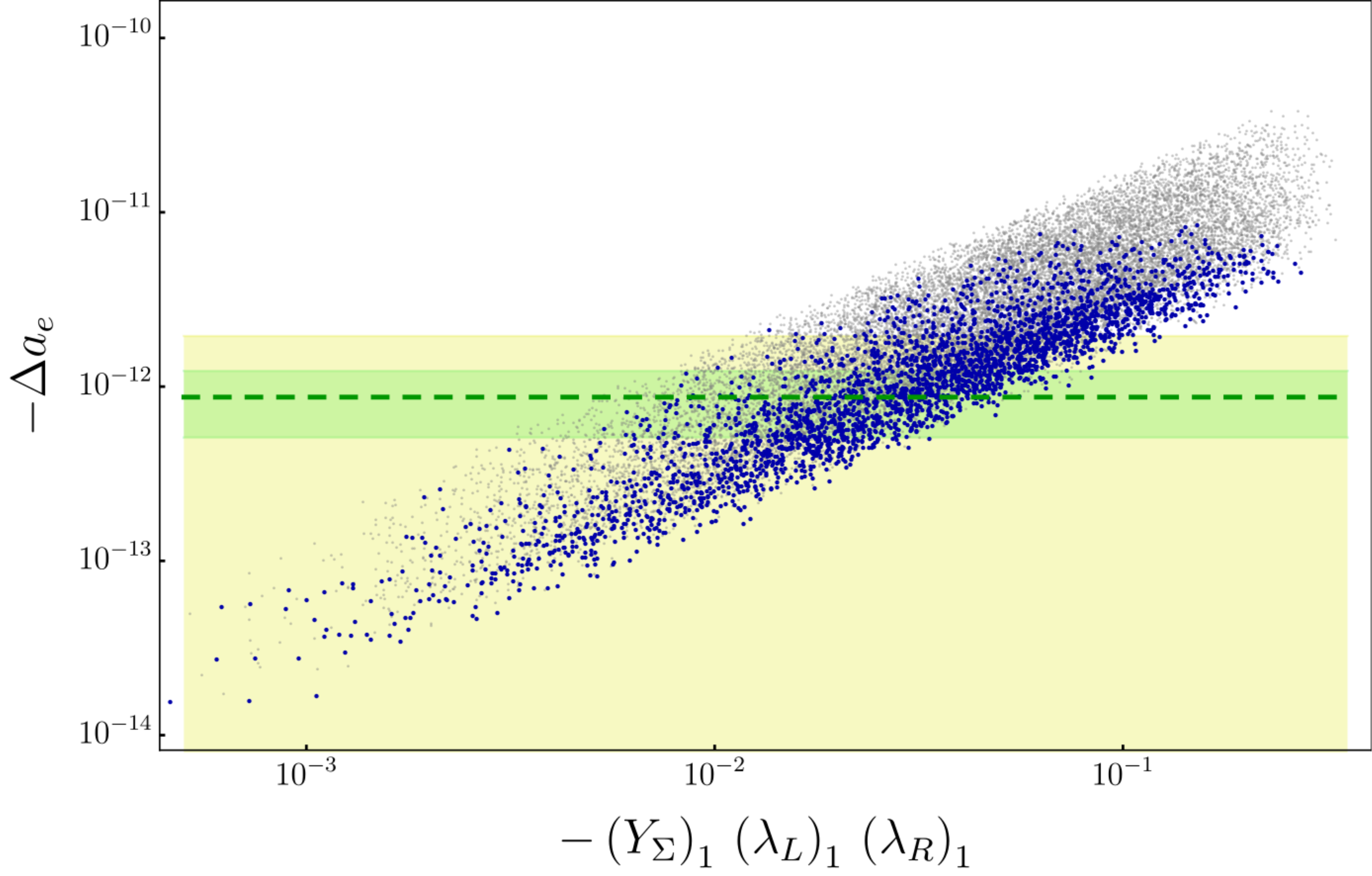}
  \end{subfigure}
  \hfill
  \begin{subfigure}{0.48\textwidth}
    \centering  
    \includegraphics[width=\linewidth]{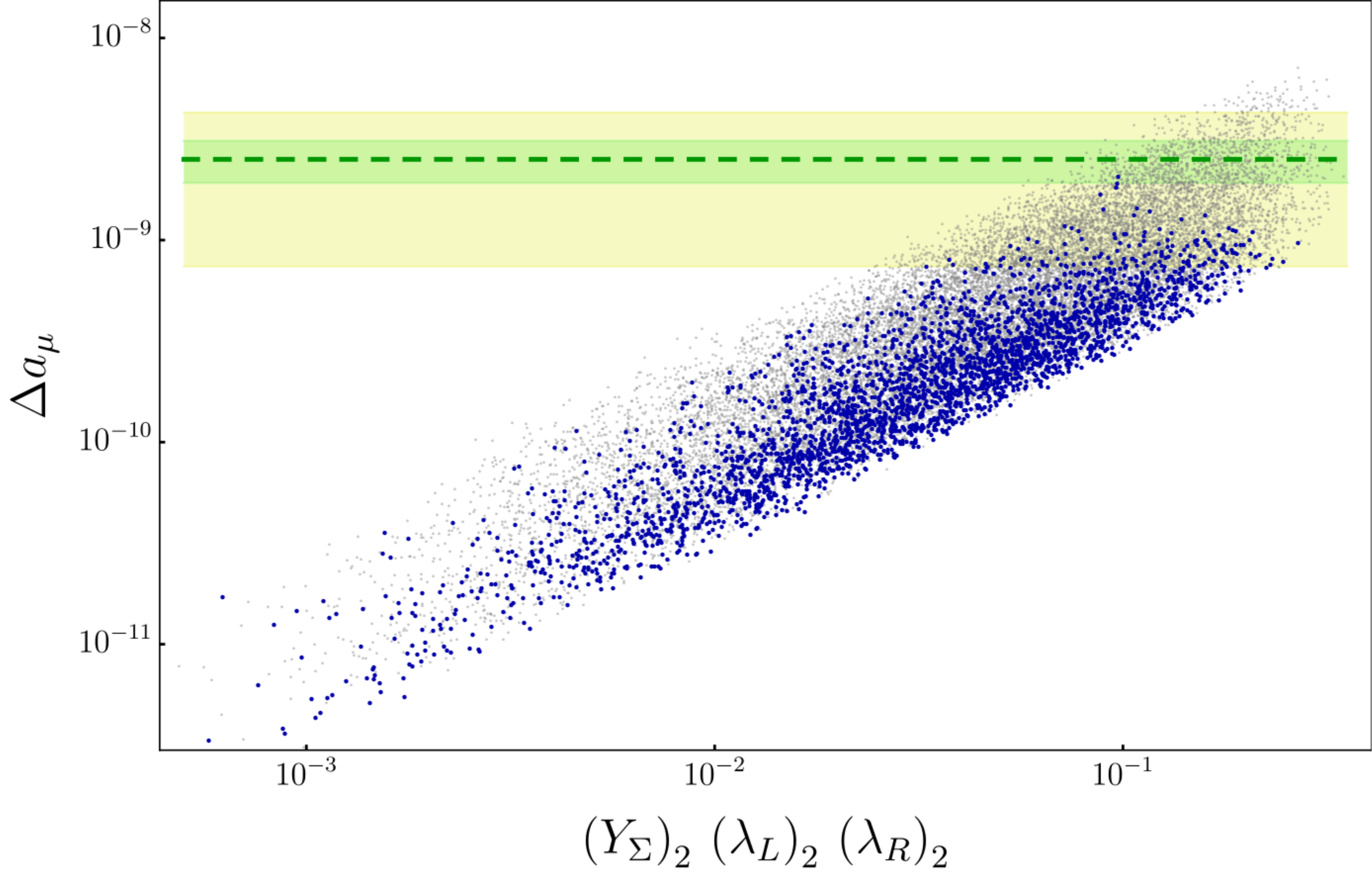}
  \end{subfigure}  
  \caption{$\Delta a_e$ (left) and $\Delta a_\mu$ (right) as a
    function of the product $\left(Y_\Sigma\right)_{ii} \,
    \left(\lambda_L\right)_i \, \left(\lambda_R\right)_i$. Blue dots
    correspond to parameter points that pass all the experimental
    constraints, whereas gray points are experimentally excluded. The
    horizontal dashed lines represent the central values for $\Delta
    a_e$ and $\Delta a_\mu$, whereas $1 \sigma$ ($3 \sigma$) regions
    are displayed as yellow (green) bands.
    \label{fig:corr}
  }
\end{figure}

\begin{figure}[!tb]
  \centering
  \includegraphics[width=0.6\linewidth]{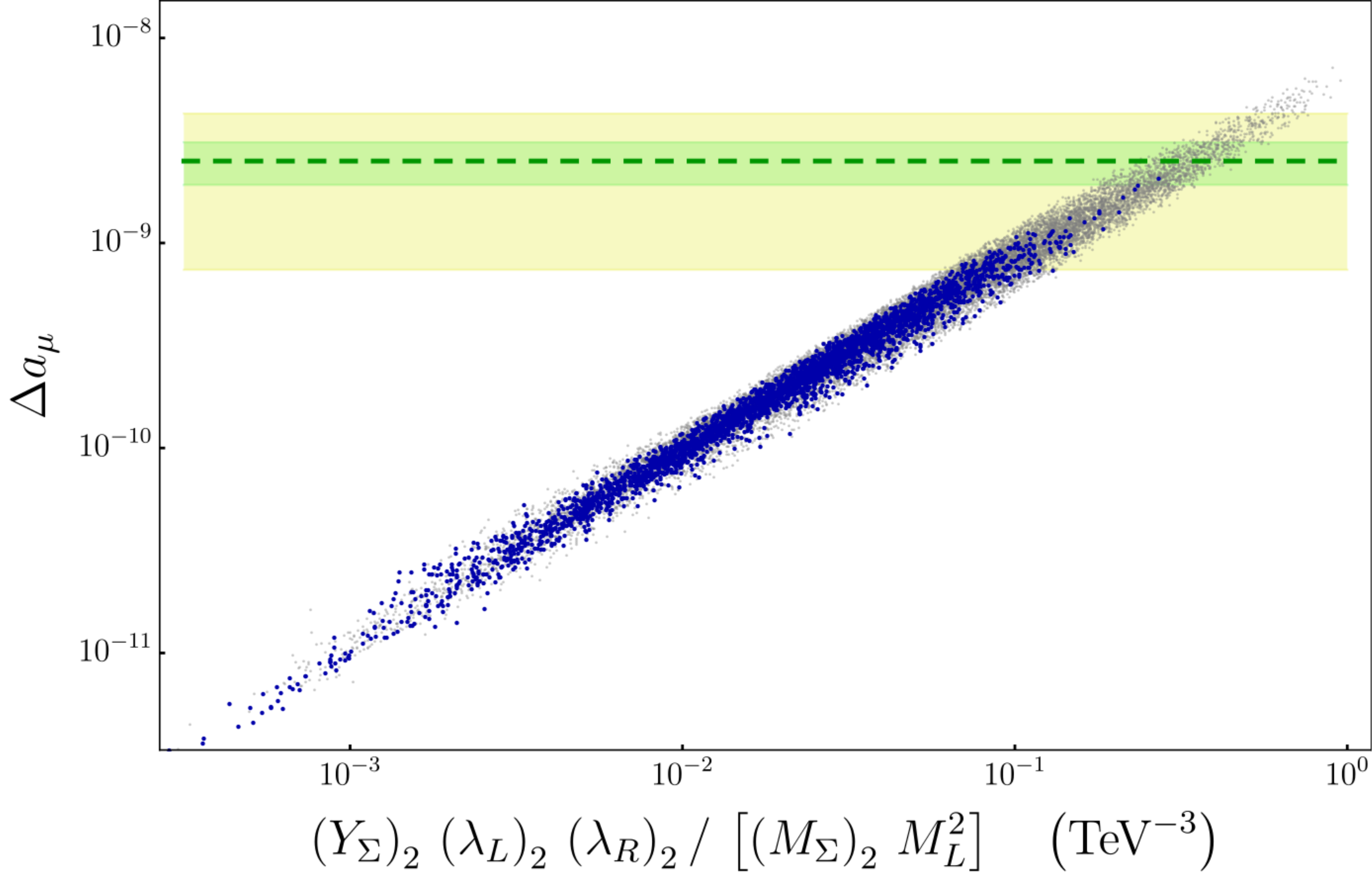}
  \caption{$\Delta a_\mu$ as a function of the combination
    $\left(Y_\Sigma\right)_{22} \, \left(\lambda_L\right)_2 \,
    \left(\lambda_R\right)_2 / \left[ \left( M_\Sigma \right)_2 \,
      M_L^2 \right]$. Dashed line, horizontal bands and color code as
    in Fig.~\ref{fig:corr}.
    \label{fig:corr2}
  }
\end{figure}

Our choice of a diagonal $Y_\Sigma$ (in the basis in which $M_\Sigma$
is diagonal too) implies that the mixing among different triplets is
typically very small. In this case, and unless there is a large mixing
with the VL neutral leptons, two of the heavy neutral mass eigenstates
are given in good approximation by the $\Sigma_1-\Sigma_1^\prime$
quasi-Dirac pair and couple mainly to electrons. Similarly, two of the
heavy neutral mass eigenstates are approximately given by the
$\Sigma_2-\Sigma_2^\prime$ quasi-Dirac pair and couple mainly to
muons. Therefore, the discussion in the previous Section implies that
one expects a strong correlation between $\Delta a_e$ and the product
$\left(Y_\Sigma\right)_{11} \, \left(\lambda_L\right)_1 \,
\left(\lambda_R\right)_1$, as well as between $\Delta a_\mu$ and the
product $\left(Y_\Sigma\right)_{22} \, \left(\lambda_L\right)_2 \,
\left(\lambda_R\right)_2$. This is clearly shown in
Fig.~\ref{fig:corr}. The left side of this figure shows $\Delta a_e$,
whereas the right side displays results for $\Delta a_\mu$, in both
cases as a function of the said parameter combinations. Here, and in
the following figures, parameter points that pass all the experimental
constraints are displayed in blue color, whereas gray points are
excluded for one or several of the reasons explained above, but
included in the figure for illustration purposes. The horizontal
dashed line represents the central value for $\Delta a_\ell$, while $1
\sigma$ ($3 \sigma$) regions are displayed as yellow (green)
bands. The first and most important result shown in this figure is
that the ISS3VL can indeed explain the electron and muon $g-2$
anomalies. In the case of the electron, one can easily find fully
valid parameter points within the $1 \sigma$ region, corresponding to
values of $-\left(Y_\Sigma\right)_{11} \, \left(\lambda_L\right)_1 \,
\left(\lambda_R\right)_1$ in the ballpark of $\sim 0.01 - 0.05$. In
fact, one can even exceed the experimental hint. In contrast, the muon
$g-2$ can only be explained within $1 \sigma$ in a narrow region of
the parameter space, with $\left(Y_\Sigma\right)_{22} \,
\left(\lambda_L\right)_2 \, \left(\lambda_R\right)_2 \sim 0.1$. This
is due to the combination of constraints that apply to our
setup. Fig.~\ref{fig:corr} also confirms the correlations with the
product $\left(Y_\Sigma\right)_{ii} \, \left(\lambda_L\right)_i \,
\left(\lambda_R\right)_i$, as we expected from the arguments given in
the previous Section. The correlation is even more pronounced in terms
of the combination $\left(Y_\Sigma\right)_{ii} \,
\left(\lambda_L\right)_i \, \left(\lambda_R\right)_i / \left[ \left(
  M_\Sigma \right)_i \, M_L^2 \right]$, as shown in
Fig.~\ref{fig:corr2} for the case of the muon $g-2$. This implies that
the Feynman diagram in Fig.~\ref{fig:WISS3VL} indeed provides one of
the dominant contributions to $\Delta a_\ell$.

An example parameter point that achieves $\Delta a_e$ and $\Delta
a_\mu$ values in the $1 \sigma$ regions indicated in
Eq.~\eqref{eq:Deltaa} is given by
\begin{equation}
  \left(M_\Sigma\right)_{ii} = 1 \, \text{TeV} \, , \quad M_L = 630 \, \text{GeV} \, ,
\end{equation}
and
\begin{equation}
  \left(Y_\Sigma\right)_{ii} = 0.117 \, , \quad \left(\lambda_L\right)_1 = -0.6 \, , \quad \left(\lambda_L\right)_2 = \sqrt{4 \pi} \, , \quad \left(\lambda_R\right)_1 = 0.1 \, , \quad \left(\lambda_R\right)_2 = 0.25 \, .
\end{equation}
We note that a large $\left(\lambda_L\right)_2$, close to the
non-perturbativity regime, is required to obtain a muon $g-2$ close to
the measured central value. While in principle this is perfectly fine,
one can relax this restriction with additional contributions to the
muon $g-2$. For instance, we expect this to happen in a non-minimal
version of our model with more than just one generation of VL leptons
or including singlet VL leptons.

\begin{figure}[!tb]
  \centering
  \begin{subfigure}{0.48\textwidth}
    \centering  
    \includegraphics[width=\linewidth]{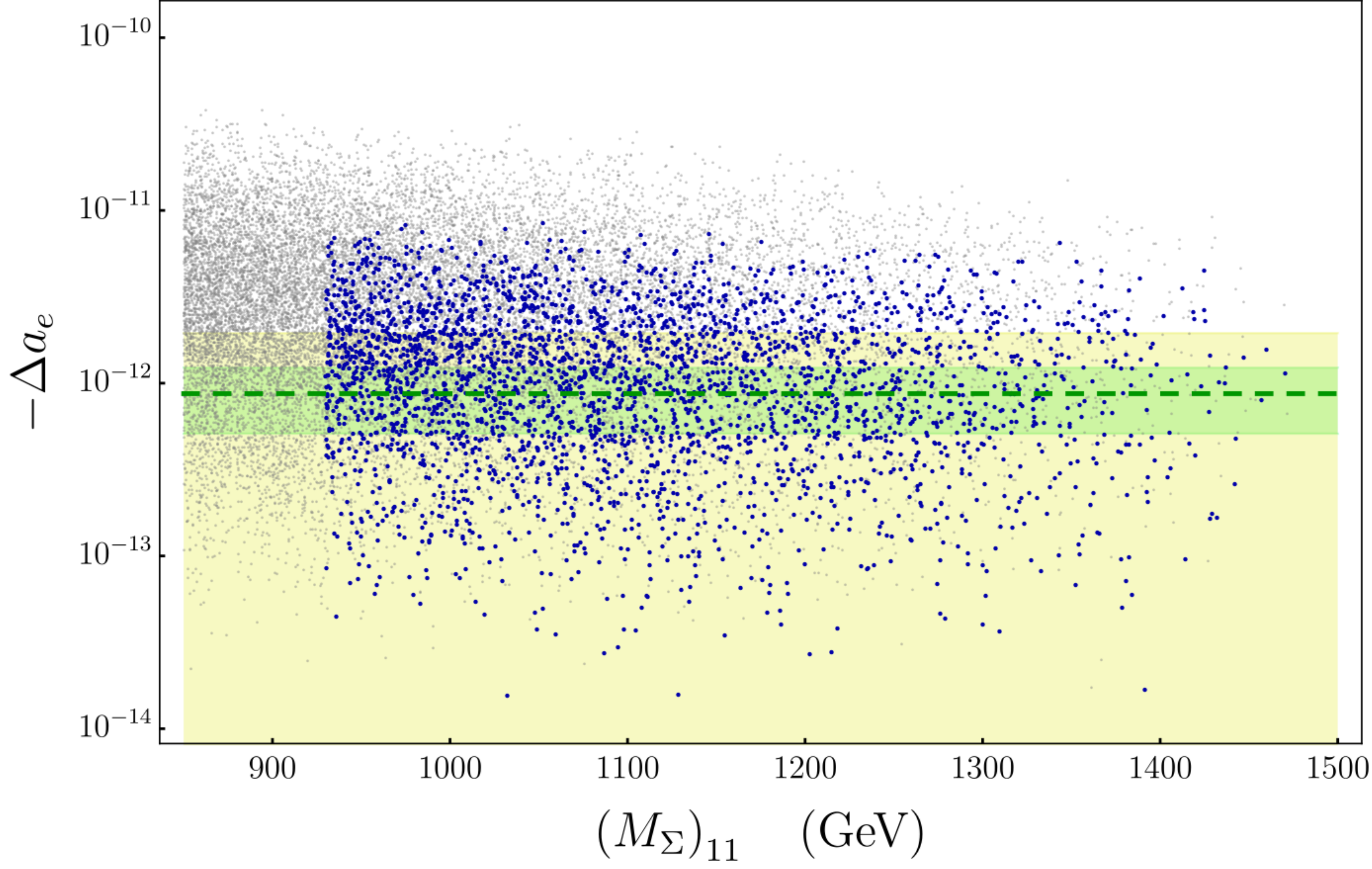}
  \end{subfigure}
  \hfill
  \begin{subfigure}{0.48\textwidth}
    \centering  
    \includegraphics[width=\linewidth]{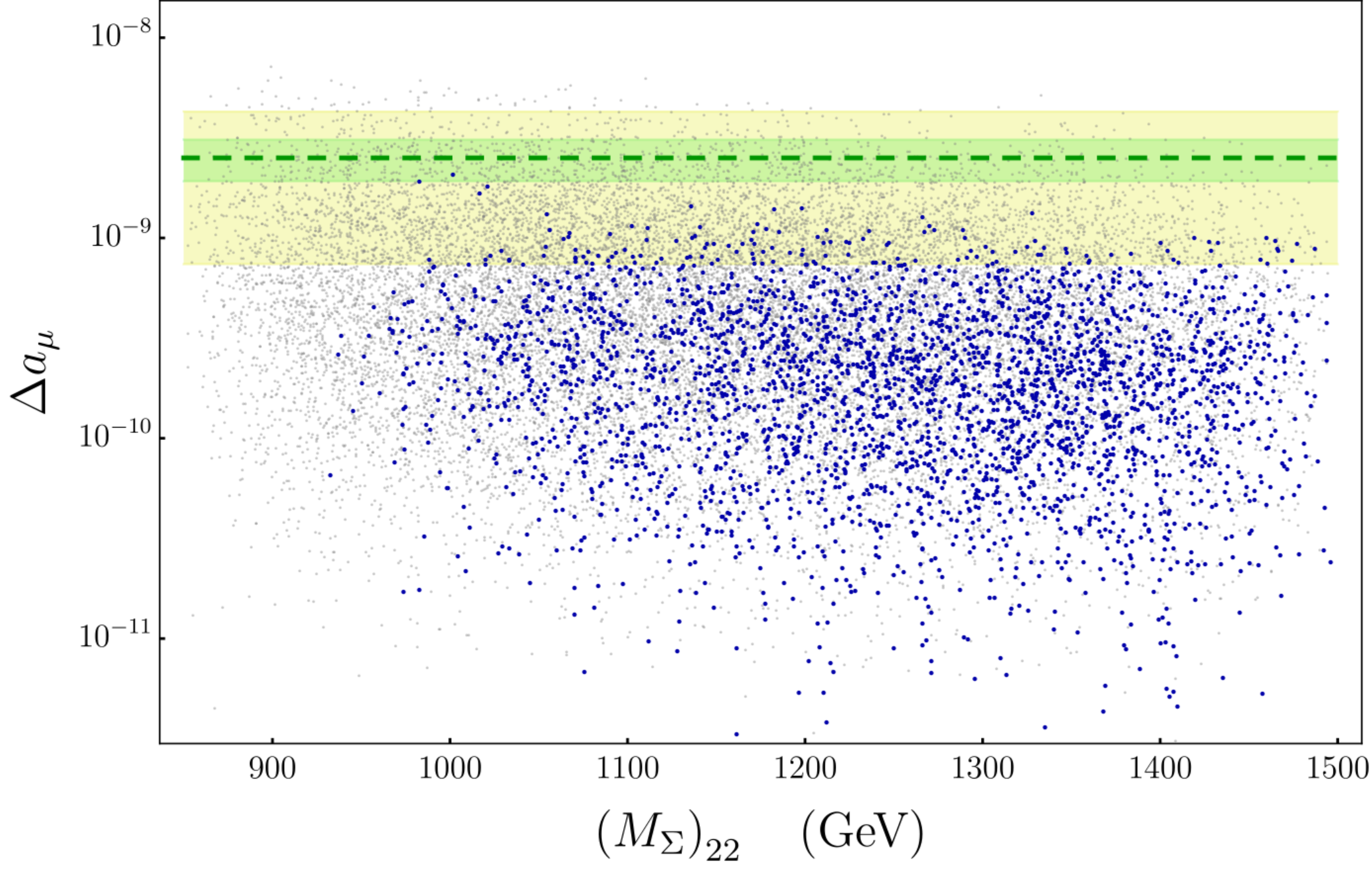}
  \end{subfigure}  
  \caption{$\Delta a_e$ (left) and $\Delta a_\mu$ (right) as a
    function of $\left(M_\Sigma\right)_{11}$ and
    $\left(M_\Sigma\right)_{22}$, respectively. Dashed line,
    horizontal bands and color code as in Fig.~\ref{fig:corr}.
    \label{fig:DeltaaVSMS}
  }
\end{figure}

\begin{figure}[!tb]
  \centering
  \includegraphics[width=0.6\linewidth]{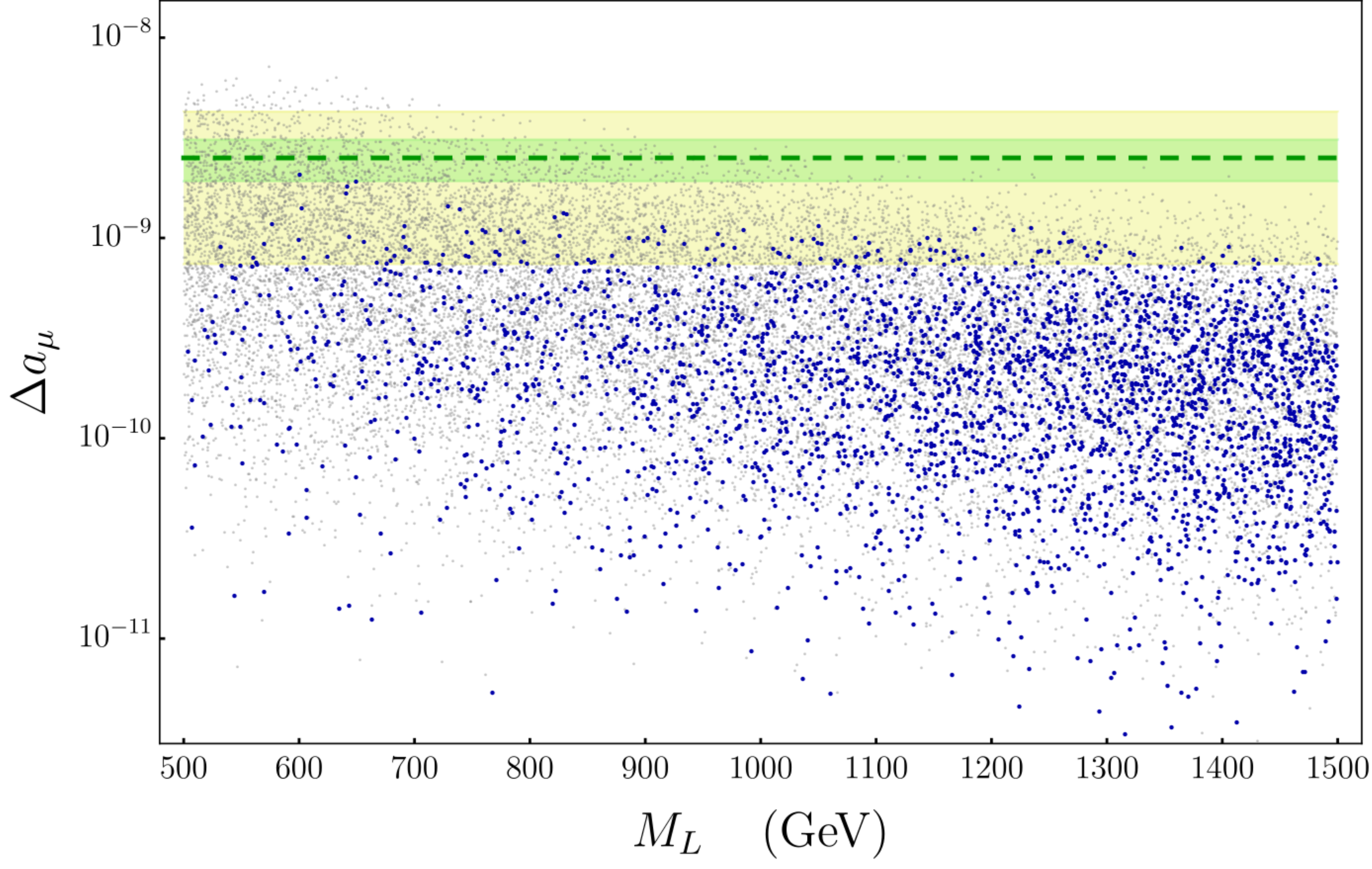}
  \caption{$\Delta a_\mu$ as a function of $M_L$. Dashed line,
    horizontal bands and color code as in Fig.~\ref{fig:corr}.
    \label{fig:DeltaaVSML}
  }
\end{figure}

Fig.~\ref{fig:DeltaaVSMS} shows the dependence of $\Delta a_\ell$ on
$\left(M_\Sigma\right)_{ii}$. On the left side, $\Delta a_e$ is shown
as a function of $\left(M_\Sigma\right)_{11}$, whereas the right side
panel shows $\Delta a_\mu$ as a function of
$\left(M_\Sigma\right)_{22}$. As explained above, these are the
parameters that determine the masses of the tiplets that couple mainly
to electrons and muons, respectively. Therefore, as expected, the NP
contributions decrease for larger values of
$\left(M_\Sigma\right)_{ii}$. However, given the limited range over
which these parameters were scanned, the reduction is not very
strong. A lower bound $\left(M_\Sigma\right)_{11} \gtrsim 930$ GeV is
clearly visible on the left-side panel. This is due to the fact that
lower $\left(M_\Sigma\right)_{11}$ values would lead to lighter
triplets, excluded by CMS searches. The dependence on $M_L$, the VL
mass, is shown in Fig.~\ref{fig:DeltaaVSML} for the case of the muon
$g-2$. One can clearly see in this plot, as well as in the previous
ones, that the density of valid parameter points gets reduced for low
masses. This is because low $\left(M_\Sigma\right)_{ii}$ and/or $M_L$
often lead to exclusion due to the $\Gamma(Z \to \ell^+ \ell^-)$
constraint.

\begin{figure}[!tb]
  \centering
  \begin{subfigure}{0.48\textwidth}
    \centering  
    \includegraphics[width=\linewidth]{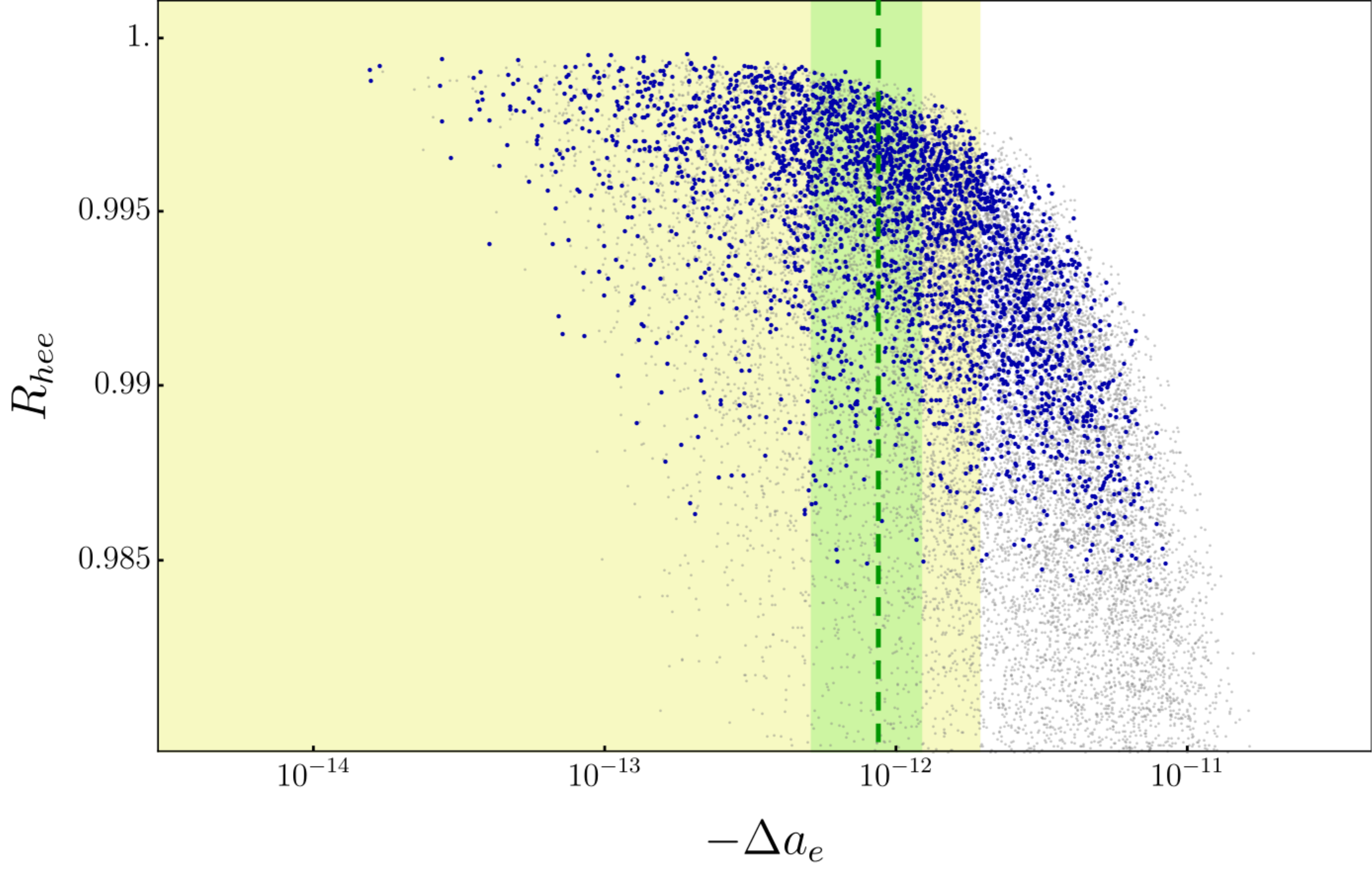}
  \end{subfigure}
  \hfill
  \begin{subfigure}{0.48\textwidth}
    \centering  
    \includegraphics[width=\linewidth]{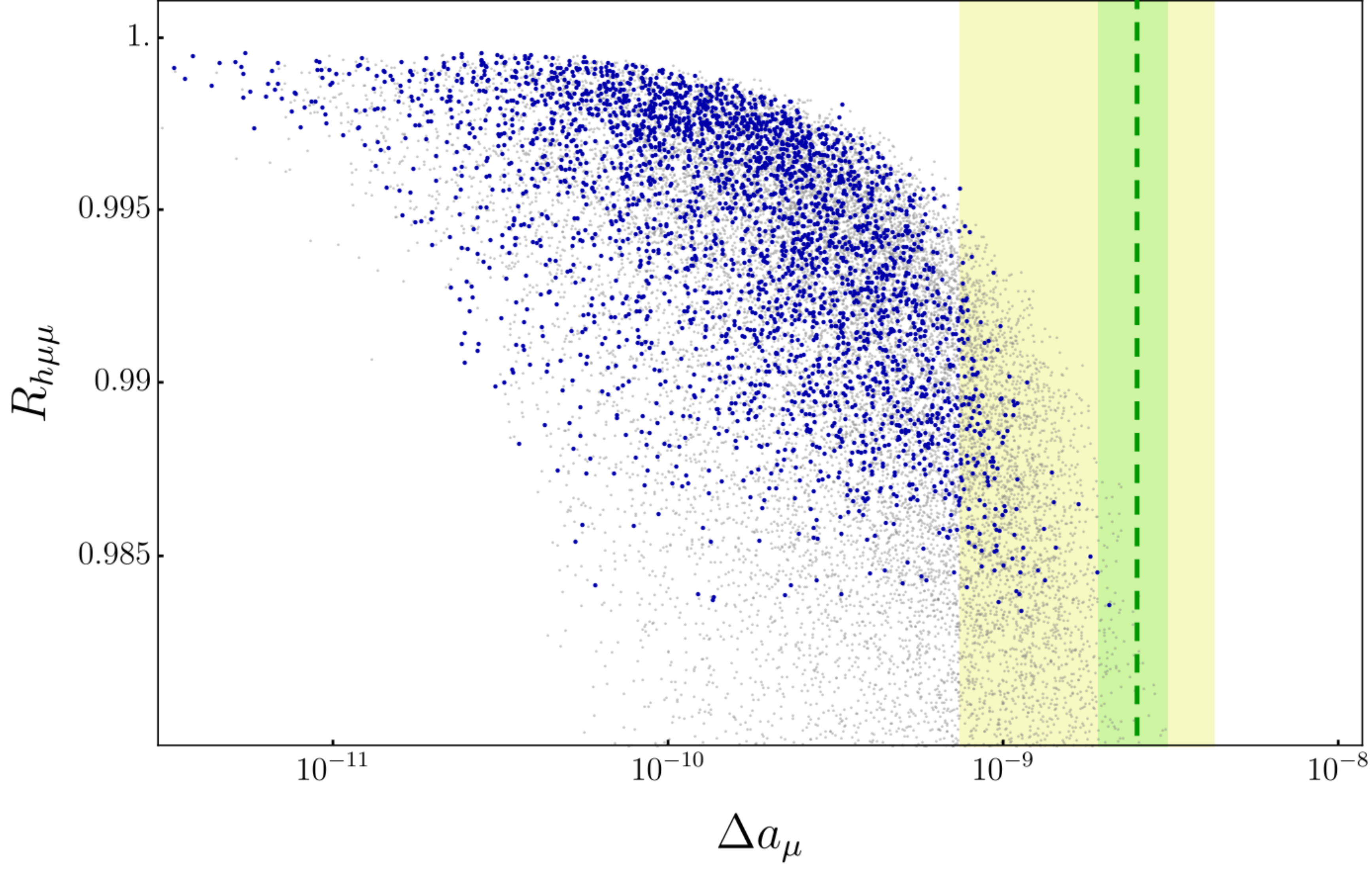}
  \end{subfigure}  
  \caption{$R_{hee}$ (left) and $R_{h\mu\mu}$ (right) as a function
    of $\Delta a_e$ and $\Delta a_\mu$, respectively.  The
    vertical dashed lines represent the central values for $\Delta
    a_e$ and $\Delta a_\mu$, whereas $1 \sigma$ ($3 \sigma$) regions
    are displayed as yellow (green) bands.
    \label{fig:Hdec}
  }
\end{figure}

We turn now our attention to the predictions of our
setup. Fig.~\ref{fig:Hdec} shows the predicted values for the ratios
\begin{equation}
  R_{h \ell \ell} = \frac{\Gamma(h \to \ell^+ \ell^-)}{\Gamma_{\rm SM}(h \to \ell^+ \ell^-)} \, ,
\end{equation}
where $\Gamma_{\rm SM}(h \to \ell^+ \ell^-)$ is the SM decay width, as
a function of $\Delta a_\ell$. In both cases, the reductions are
small, at most of about $\sim 2\%$ in the parameter points that pass
all the experimental bounds. This is due the fact that the
$R_{h\ell\ell}$ ratios correlate with the corresponding
$R_{Z\ell\ell}$ ratios, and are thus strongly constrained. Therefore,
our setup predicts SM-like Higgs boson decays into charged leptons.

Finally, we have focused on a scenario with diagonal $Y_\Sigma$
couplings in order to enhace the diagonal contributions to the muon
$g-2$. This strongly suppresses all lepton flavor violating signals,
which can only take place via the off-diagonal entries of the small
$\mu$ parameter. We believe this to be a generic prediction of our
model when the muon $g-2$ anomaly is addressed. However, we cannot
discard the possibility of very fine-tuned parameter regions with
large off-diagonal $Y_\Sigma$ couplings and accidentally suppressed
flavor violating transitions.

\section{Discussion}
\label{sec:conclusions}

The recent announcement of the first results by the Muon $g-2$
collaboration at Fermilab has sparked a renewed interest in a
long-standing anomaly. Together with the analogous discrepancy in the
anomalous magnetic moment of the electron, they constitute a pair of
intriguing deviations with respect to the SM predictions. If
confirmed, new BSM states with masses not much above the electroweak
scale will be required in order to address the discrepancies.

We have analyzed the electron and muon $g-2$ in an extended version of
the ISS3 model that includes a pair of VL doublet leptons. This model
is motivated by the need to generate neutrino masses, which in this
setup are induced at the electroweak scale. This naturally leads to a
rich phenomenology in multiple fronts. Our analysis has taken into
account the most relevant experimental bounds in our scenario. This
includes limits from direct searches at the LHC, deviations in $Z \to
\ell^+ \ell^-$ decays and a compilation of electroweak limits. These
are the main conclusions of our work:
\begin{itemize}
  \item The pure ISS3 cannot address the electron and muon $g-2$
    anomalies due to the combination of the constraints on the
    $m_D/M_\Sigma$ ratio derived from a variety of electroweak data
    and a strong chiral suppression of the order of $m_\ell/v$. In
    addition, the contributions to the muon $g-2$ have the wrong sign.
  \item The inclusion of a VL lepton doublet pair to the ISS3 particle
    content suffices to enhance the contributions to the muon $g-2$,
    allows one to adjust its sign conveniently and fully addresses the
    observed discrepancy.
  \item The electron $g-2$ anomaly can also be explained, in this
    case in a wider region of the parameter space.
  \item No significant change in the $h \to \ell^+ \ell^-$ decays is
    found, and then these stay SM-like.
\end{itemize}

In this work we have focused on a minimal extension of the ISS3,
containing only one VL lepton doublet pair. In principle, one may
introduce additional copies of $L_{L,R}$ or VL lepton singlets, see
for instance~\cite{Dermisek:2013gta}. These non-minimal variations may
reduce the impact of some of the bounds or enlarge the parameter space
compatible with the current experimental hints in the muon
$g-2$. Furthermore, lepton number is explicitly broken in our
model. Alternatively, one may consider the spontaneous breaking of a
global $\rm U(1)_L$ lepton number symmetry, leading to the appearance
of a massless Goldstone boson, the majoron. However, a pure massless
pseudoscalar state gives negative (lepton flavor conserving)
contributions to the electron and muon anomalous magnetic
moments~\cite{Escribano:2020wua}, hence being unable to solve the
tension in the case of the muon $g-2$.

Exciting times are ahead of us. The muon $g-2$ anomaly, now hinted by
a second experiment, joins the list of results that have recently
attracted attention to muons. It is natural to speculate about this
anomaly together with the $R_K$ and $R_{K^*}$ anomalies found by the
LHCb collaboration, as well as with the set of deviations observed in
recent years in semileptonic $b \to s$ and $b \to c$ transitions. New
experimental results, that may finally confirm an emerging picture
beyond the SM, are eagerly awaited.

\section*{Acknowledgements}

The authors are grateful to Martin Hirsch, Farinaldo Queiroz and
Moritz Platscher for fruitful discussions. They also would like to
thank Ricardo Cepedello for help with the code and Saiyad Ashanujjaman
for pointing out early references discussing the charged leptons $g-2$
in seesaw scenarios. Work supported by the Spanish grants
FPA2017-85216-P (MINECO/AEI/FEDER, UE), SEJI/2018/033 (Generalitat
Valenciana) and FPA2017-90566-REDC (Red Consolider MultiDark). The
work of PE is supported by the FPI grant PRE2018-084599. AV
acknowledges financial support from MINECO through the Ramón y Cajal
contract RYC2018-025795-I. The work of JTC is supported by the
Ministerio de Ciencia e Innovaci\'on under FPI contract PRE2019-089992
of the SEV-2015-0548 grant and the grant PGC2018-102016-A-I00.

\appendix

\section{Couplings}
\label{sec:app1}

The couplings involved in the calculation of the charged lepton
anomalous magnetic moments are shown in the Feynman diagrams of
Fig.~\ref{fig:loops} and have been computed with the help of {\tt
  SARAH}~\cite{Staub:2008uz,Staub:2009bi,Staub:2010jh,Staub:2012pb,Staub:2013tta}.~\footnote{See~\cite{Vicente:2015zba}
  for a pedagogical introduction to the use of {\tt SARAH}.}. We
define them and list their analytical expressions here:


\begin{center} 
  \begin{fmffile}{FeynDia57} 
  \fmfframe(20,20)(20,20){ 
  \begin{fmfgraph*}(75,75) 
  \fmfleft{l1}
  \fmfright{r1,r2}
  \fmf{plain}{l1,v1}
  \fmf{fermion}{r1,v1}
  \fmf{boson}{r2,v1}
  \fmflabel{$N_{{i}}$}{l1}
  \fmflabel{$\chi_{{j}}$}{r1}
  \fmflabel{$W^+_{{\mu}}$}{r2}
  \end{fmfgraph*}} 
  \end{fmffile} 
  \end{center}

\begin{equation}
\mathcal{L}_{\chi N W} = \bar{N}_i \, \gamma^\mu \, \left[\left(g^L_{\chi N W}\right)_{ij} P_L+\left(g^R_{\chi NW}\right)_{ij} P_R\right] \chi_j \, W_\mu + \hc
\end{equation}

\begin{align} 
  &\left(g^L_{\chi N W}\right)_{ij} = - g \left( \frac{1}{\sqrt{2}} \sum_{a=1}^3 \mathcal{V}^{L \, *}_{j a} \, \mathcal{U}_{i a} + \sum_{a=4}^{9} \mathcal{V}^{L \, *}_{j a} \, \mathcal{U}_{i a} + \frac{1}{\sqrt{2}} \, \mathcal{V}^{L \, *}_{j 10} \, \mathcal{U}_{i 10} \right) \\ 
  &\left(g^R_{\chi N W}\right)_{ij} = - g \left( \sum_{a=4}^9 \mathcal{V}^R_{j a} \, \mathcal{U}^*_{i a} - \frac{1}{\sqrt{2}} \, \mathcal{V}^R_{j 10} \, \mathcal{U}^*_{i 11} \right) 
\end{align} 


\begin{center} 
  \begin{fmffile}{FeynDia59} 
  \fmfframe(20,20)(20,20){ 
  \begin{fmfgraph*}(75,75) 
  \fmfleft{l1}
  \fmfright{r1,r2}
  \fmf{fermion}{v1,l1}
  \fmf{fermion}{r1,v1}
  \fmf{wiggly}{r2,v1}
  \fmflabel{$\bar{\chi}_{{i}}$}{l1}
  \fmflabel{$\chi_{{j}}$}{r1}
  \fmflabel{$Z_{{\mu}}$}{r2}
  \end{fmfgraph*}} 
  \end{fmffile} 
\end{center}

\begin{equation}
\mathcal{L}_{\chi Z} = \bar{\chi}_i \, \gamma^\mu \, \left[\left(g^L_{\chi Z}\right)_{ij} P_L+\left(g^R_{\chi Z}\right)_{ij} P_R\right] \chi_j \, Z_\mu + \hc
\end{equation}

\begin{align} 
  & \left(g^L_{\chi Z}\right)_{ij} = \frac{1}{2} \left(g \cos \theta_W - g' \sin \theta_W \right) \left( \sum_{a=1}^3 \mathcal{V}^{L \, *}_{j a} \, \mathcal{V}^L_{i a} + \mathcal{V}^{L \, *}_{j 10} \, \mathcal{V}^L_{i 10} \right) + g \cos \theta_W  \sum_{a=4}^9 \mathcal{V}^{L \, *}_{j a} \, \mathcal{V}^L_{i a} \\ 
  & \left(g^R_{\chi Z}\right)_{ij} = - g' \sin \theta_W \sum_{a=1}^3 \mathcal{V}^{R \, *}_{i a} \, \mathcal{V}^R_{j a} + g \cos \theta_W \sum_{a=4}^9 \mathcal{V}^{R \, *}_{i a} \, \mathcal{V}^R_{j a} + \frac{1}{2} \left(g \cos \theta_W - g' \sin \theta_W \right) \, \mathcal{V}^{R \, *}_{i 10} \, \mathcal{V}^R_{j 10}
\end{align}

        
\begin{center} 
  \begin{fmffile}{FeynDia48} 
    \fmfframe(20,20)(20,20){ 
      \begin{fmfgraph*}(75,75) 
        \fmfleft{l1}
        \fmfright{r1,r2}
        \fmf{fermion}{v1,l1}
        \fmf{fermion}{r1,v1}
        \fmf{dashes}{r2,v1}
        \fmflabel{$\bar{\chi}_{{i}}$}{l1}
        \fmflabel{$\chi_{{j}}$}{r1}
        \fmflabel{$h$}{r2}
      \end{fmfgraph*}} 
  \end{fmffile} 
\end{center}

\begin{equation}
\mathcal{L}_{\chi h} = \bar{\chi}_i \, \left[ \left(g^L_{\chi h}\right)_{ij} P_L+\left(g^R_{\chi h}\right)_{ij} P_R \right] \chi_j \, h + \hc
\end{equation}

\begin{align} 
  &\left( g^L_{\chi h} \right)_{i j} = -\frac{1}{\sqrt{2}} \sum_{a,b=1}^{3} \mathcal{V}^{L \, *}_{j b} \, \mathcal{V}^{R \, *}_{i a} \, \left(Y_e\right)_{a b} - \sum_{a=1}^3 \sum_{b=1}^{3} \mathcal{V}^{L \, *}_{j b} \, \mathcal{V}^{R \, *}_{i \, a+3} \, \left(Y_\Sigma\right)_{a b} \nonumber \\
  &\hspace{1.75cm} - \mathcal{V}^{L \, *}_{j 10} \left[ \sum^3_{a=1} \mathcal{V}^{R \, *}_{i \, a+3} \, \left(\lambda_L\right)_a + \frac{1}{\sqrt{2}} \sum^3_{a=1} \mathcal{V}^{R \, *}_{i a} \, \left(\lambda_R\right)_a \right] \\
  & \left( g^R_{\chi h} \right)_{i j} = -\frac{1}{\sqrt{2}} \sum_{a,b=1}^{3} \mathcal{V}^L_{i b} \, \mathcal{V}^R_{j a} \, \left(Y_e\right)^*_{a b} - \sum_{a=1}^3 \sum_{b=1}^{3} \mathcal{V}^L_{i b} \, \mathcal{V}^R_{j \, a+3} \, \left(Y_\Sigma\right)^*_{a b} \nonumber \\
  &\hspace{1.75cm} - \left[ \sum^3_{a=1} \mathcal{V}^{R}_{j \, a+3} \, \left(\lambda_L^*\right)_a + \frac{1}{\sqrt{2}} \sum^3_{a=1} \mathcal{V}^{R}_{j a} \, \left(\lambda_R^*\right)_a \right] \, \mathcal{V}^{L}_{i 10} 
\end{align} 
   

\begin{center} 
  \begin{fmffile}{FeynDia58} 
    \fmfframe(20,20)(20,20){ 
      \begin{fmfgraph*}(75,75) 
          \fmfleft{l1}
          \fmfright{r1,r2}
          \fmf{fermion}{v1,l1}
          \fmf{fermion}{r1,v1}
          \fmf{wiggly}{r2,v1}
          \fmflabel{$\bar{\chi}_{{i}}$}{l1}
          \fmflabel{$\chi_{{j}}$}{r1}
          \fmflabel{$\gamma_{{\mu}}$}{r2}
      \end{fmfgraph*}} 
  \end{fmffile} 
 \end{center}  
\begin{equation}
  \mathcal{L}_{\chi \gamma} = \bar{\chi}_i \, \gamma^\mu \, \left[\left(g^L_{\chi \gamma}\right)_{ij} P_L+\left(g^R_{\chi \gamma}\right)_{ij} P_R\right] \chi_j \, A_\mu + \hc
\end{equation}

  \begin{equation} 
    \left(g^L_{\chi \gamma}\right)_{ij} = \left(g^R_{\chi \gamma}\right)_{ij} = - e \, \delta_{ij}
  \end{equation} 

  
\begin{center} 
  \begin{fmffile}{FeynDia39} 
   \fmfframe(20,20)(20,20){ 
    \begin{fmfgraph*}(75,75) 
      \fmfleft{l1}
      \fmfright{r1,r2}
      \fmf{boson}{v1,l1}
      \fmf{wiggly}{r1,v1}
      \fmf{boson}{r2,v1}
      \fmflabel{$W^-_{1}$}{l1}
      \fmflabel{$\gamma$}{r1}
      \fmflabel{$W^+_{2}$}{r2}
    \end{fmfgraph*}} 
  \end{fmffile} 
\end{center}  
   
\begin{equation}
  \mathcal{L}_{W \gamma} =\Gamma^{\mu\alpha\beta} \, W_\alpha W_\beta A_\mu
\end{equation}

  \begin{align}
    &\Gamma^{\mu\alpha\beta} = g \, \sin\theta_W \left[g^{\alpha\beta}\left(p_{W_1}^\mu+p_{W_2}^\mu\right)+
    g^{\mu\beta}\left(-p_{W_2}^\alpha-p_{\gamma}^\alpha\right)+g^{\alpha\mu}\left(p_{\gamma}^\beta-p_{W1}^\beta\right)\right]
  \end{align}

\section{Charged lepton anomalous magnetic moments: full expressions}
\label{sec:app2}

We define the dimensionless quantities
\begin{equation}
  \epsilon_{\ell i} = \dfrac{m_\ell}{m_{\chi_i}} \, , \quad \delta_{\ell i} =\dfrac{m_\ell}{m_{N_i}} \, , \quad \omega_{ai} = \dfrac{m_a}{m_{\chi_i}} \quad \text{and} \quad \omega_{W i} = \dfrac{m_W}{m_{N_i}} \, ,
\end{equation}
with $a=Z,\, h$.

\subsubsection*{$\boldsymbol{W}$ contribution}

The $W$ contribution to the charged lepton anomalous magnetic moments
can be written as
\begin{equation}
  \Delta a_{\ell} \left( W \right) =  \dfrac{1}{32 \pi^2 \, \omega_{W i}^2 \delta_{\ell i}^4} \left[ \left(C^2_{\chi N W}\right)_{i \ell} \, f_W(\delta_{\ell i}^2,\omega_{W i}^2) - \delta_{\ell i} \, \left(D^2_{\chi N W}\right)_{i \ell} \, g_W(\delta_{\ell i}^2,\omega_{W i}^2) \right] \, ,
\end{equation}
where a sum over the repeated index $i$ is implicit and $f_W$ and $g_W$ are two loop functions given by
\begin{align}
  f_W(x,y) &= x^3 + x^2 \left(8\, y - 1\right) + 2\, x\, \left(1+y-2 \, y^2\right) \nonumber \\
  & + \left[3 \, x^2 \, y - x \left( 1 - 3 \, y + 5 \, y^2\right) - 3 \, y^2 + 2 \, y^3 + 1 \right] \log y \nonumber \\
  & + \frac{2}{\Delta(x,y)} \left[3\, x^3 \, y + x^2 \left(1 - 8\,y^2\right) + x \left(7\, y^3 - 7\, y^2 + 2\, y - 2\, \right) \right. \nonumber \\
    & \left. + \left(1-y\right)^3 \left( 1 + 2\, y \right)  \right] \log \frac{1+y-x + \Delta(x,y)}{2 \, \sqrt{y}} \, , \\
  g_W(x,y) &= 2 \, x \left( 1+ 2 \, y\right) +  \, \left[ x \, \left( 3 \, y - 1 \right) - 2 \, y^2 \, + y + 1 \right] \, \log y  \nonumber \\
  & + \frac{2}{\Delta(x,y)} \left[ x^2 \left( 3 \, y +1 \right) - x \left(2 - y + 5 \, y^2 \right) + 2 \, y^3 - 3 \, y^2 + 1 \right] \log \frac{1+y-x + \Delta(x,y)}{2 \, \sqrt{y}} \, .
\end{align}
Here we have defined the auxiliary function
\begin{equation}
  \Delta(x,y) = \sqrt{x^2 - 2 \, x \left(1+y\right) + \left(1-y\right)^2} \, .
\end{equation}

\subsubsection*{$\boldsymbol{Z}$ contribution}

The $Z$ contribution can be written as
\begin{equation}
  \Delta a_{\ell} \left( Z \right) = \dfrac{1}{32 \pi^2 \, \omega_{Z i}^2 \epsilon_{\ell i}^4} \left[ \left( C_{\chi Z}^2 \right)_{i \ell} \, f_Z(\epsilon_{\ell i}^2,\omega_{Z i}^2) + \epsilon_{\ell i} \, \left( D_{\chi Z}^2 \right)_{i \ell} \, g_Z(\epsilon_{\ell i}^2,\omega_{Z i}^2) \right] \, .
\end{equation}
Again, there is a sum over the repeated index $i$, whereas $f_Z$ and
$g_Z$ are the loop functions
\begin{align}
  f_Z(x,y) &= - x \left[ x^2 + x \left(3 - 4 \, y \right) + 4 \, y^2 - 2 \, y -2 \right] \nonumber\\
  & + \left[ x^2 - x \left( 2 - 2 \, y + 3 \, y^2 \right) + 2 \, y^3 - 3 \, y^2 + 1 \right] \log y \nonumber\\
  & - \frac{2}{\Delta(x,y)}\left[x^3 + x^2 \left(3 \, y^2 + y - 3\right) - x \left(5 \, y^3 - 4 \, y^2 + 2 \, y - 3\right)  \right.\nonumber\\
    & \left. - \left(1-y\right)^3 \left(1+ 2 \, y\right)\right] \log \frac{1+y-x+\Delta(x,y)}{2 \, \sqrt{y}} \, , \\
  g_Z(x,y) &= 2 \, x \left( 2 \, x - 2 \, y - 1 \right) - \, \left[ x^2 + x \left( y - 2 \right)  - 2 \, y^2 + y +1 \right] \log y \nonumber \\
  & + \frac{2}{\Delta(x,y)} \left[ x^3 - 3 \, x^2 + 3 \, x \left( 1+y^2\right) - 2 \, y^3 + 3 \, y^2 -1 \right] \log \frac{1+y-x + \Delta(x,y)}{2 \, \sqrt{y} } \, .
\end{align}

\subsubsection*{$\boldsymbol{h}$ contribution}

Finally, the $h$ contribution to the charged leptons $g-2$ can be written as
\begin{equation}
  \Delta a_{\ell} \left( h \right) = \dfrac{1}{32 \pi^2 \, \epsilon_{\ell i}^4} \left[ \left(C_{\chi h}^2\right)_{i \ell} \, f_h(\epsilon_{\ell i}^2,\omega_{h i}^2) + \epsilon_{\ell i} \, \left ( D_{\chi h}^2 \right)_{i \ell} \, g_h(\epsilon_{\ell i}^2,\omega_{h i}^2) \right] \, ,
\end{equation}
with an implicit sum over the repeated $i$ index and the loop functions
\begin{align}
  f_h(x,y) &= -x \left( x + 2 \, y - 2 \right) + \left[ \left( 1-y \right)^2 - x \right] \log y \nonumber \\
  & + \dfrac{2}{\Delta(x,y)} \left[x^2 + x \left( y^2 + y - 2\right) + \left( 1-y\right)^3 \right] \log \dfrac{1+y-x+\Delta(x,y)}{2 \, \sqrt{y}} \, , \\
  g_h(x,y) &= 2 \, x  -  \left( x+y-1 \right) \log y \nonumber \\
  & + \dfrac{2}{\Delta(x,y)} \left[ \left(1-x\right)^2 + y^2 - 2 \, y \right] \log \dfrac{1+y-x+\Delta(x,y)}{2 \, \sqrt{y}} \, .
\end{align}

We have compared our results to the general expressions provided
in~\cite{Crivellin:2018qmi}, finding full agreement. Our results also
match those recently presented in~\cite{Dermisek:2021ajd}, where a
model with similar contributions to the muon $g-2$ was
considered. Finally, analytical expressions for the contributions to
the muon $g-2$ in the limit of heavy mediators are provided
in~\cite{Lindner:2016bgg}. While we do not consider this limit in our
paper (since it would correspond to $m_W, m_Z, m_h \gg m_{N_i},
m_{\chi_i}$), it can be used to crosscheck our results. We find full
agreement.

\providecommand{\href}[2]{#2}\begingroup\raggedright\endgroup


\begin{thebibliography}{10}

  \bibitem{Aoyama:2012wj}
T.~Aoyama, M.~Hayakawa, T.~Kinoshita, and M.~Nio, ``{Tenth-Order QED
  Contribution to the Electron g-2 and an Improved Value of the Fine Structure
  Constant},'' \href{http://dx.doi.org/10.1103/PhysRevLett.109.111807}{{\em
  Phys. Rev. Lett.} {\bfseries 109} (2012) 111807},
  \href{http://arxiv.org/abs/1205.5368}{{\ttfamily arXiv:1205.5368 [hep-ph]}}.

\bibitem{Aoyama:2012wk}
T.~Aoyama, M.~Hayakawa, T.~Kinoshita, and M.~Nio, ``{Complete Tenth-Order QED
  Contribution to the Muon g-2},''
  \href{http://dx.doi.org/10.1103/PhysRevLett.109.111808}{{\em Phys. Rev.
  Lett.} {\bfseries 109} (2012) 111808},
  \href{http://arxiv.org/abs/1205.5370}{{\ttfamily arXiv:1205.5370 [hep-ph]}}.

\bibitem{Laporta:2017okg}
S.~Laporta, ``{High-precision calculation of the 4-loop contribution to the
  electron g-2 in QED},''
  \href{http://dx.doi.org/10.1016/j.physletb.2017.06.056}{{\em Phys. Lett. B}
  {\bfseries 772} (2017) 232--238},
  \href{http://arxiv.org/abs/1704.06996}{{\ttfamily arXiv:1704.06996
  [hep-ph]}}.

\bibitem{Aoyama:2017uqe}
T.~Aoyama, T.~Kinoshita, and M.~Nio, ``{Revised and Improved Value of the QED
  Tenth-Order Electron Anomalous Magnetic Moment},''
  \href{http://dx.doi.org/10.1103/PhysRevD.97.036001}{{\em Phys. Rev. D}
  {\bfseries 97} no.~3, (2018) 036001},
  \href{http://arxiv.org/abs/1712.06060}{{\ttfamily arXiv:1712.06060
  [hep-ph]}}.

\bibitem{Bennett:2006fi}
{\bfseries Muon g-2} Collaboration, G.~Bennett {\em et~al.}, ``{Final Report of
  the Muon E821 Anomalous Magnetic Moment Measurement at BNL},''
  \href{http://dx.doi.org/10.1103/PhysRevD.73.072003}{{\em Phys. Rev. D}
  {\bfseries 73} (2006) 072003},
  \href{http://arxiv.org/abs/hep-ex/0602035}{{\ttfamily arXiv:hep-ex/0602035}}.

\bibitem{Jegerlehner:2009ry}
F.~Jegerlehner and A.~Nyffeler, ``{The Muon g-2},''
  \href{http://dx.doi.org/10.1016/j.physrep.2009.04.003}{{\em Phys. Rept.}
  {\bfseries 477} (2009) 1--110},
  \href{http://arxiv.org/abs/0902.3360}{{\ttfamily arXiv:0902.3360 [hep-ph]}}.

\bibitem{Blum:2018mom}
{\bfseries RBC, UKQCD} Collaboration, T.~Blum, P.~Boyle, V.~Gülpers,
  T.~Izubuchi, L.~Jin, C.~Jung, A.~Jüttner, C.~Lehner, A.~Portelli, and
  J.~Tsang, ``{Calculation of the hadronic vacuum polarization contribution to
  the muon anomalous magnetic moment},''
  \href{http://dx.doi.org/10.1103/PhysRevLett.121.022003}{{\em Phys. Rev.
  Lett.} {\bfseries 121} no.~2, (2018) 022003},
  \href{http://arxiv.org/abs/1801.07224}{{\ttfamily arXiv:1801.07224
  [hep-lat]}}.

\bibitem{PhysRevLett.126.141801}
{\bfseries Muon $\boldsymbol{g\ensuremath{-}2}$} Collaboration, B.~Abi {\em
  et~al.}, ``Measurement of the positive muon anomalous magnetic moment to 0.46
  ppm,'' \href{http://dx.doi.org/10.1103/PhysRevLett.126.141801}{{\em Phys.
  Rev. Lett.} {\bfseries 126} (2021) 141801}.

\bibitem{Aoyama:2020ynm}
T.~Aoyama {\em et~al.}, ``{The anomalous magnetic moment of the muon in the
  Standard Model},''
  \href{http://dx.doi.org/10.1016/j.physrep.2020.07.006}{{\em Phys. Rept.}
  {\bfseries 887} (2020) 1--166},
  \href{http://arxiv.org/abs/2006.04822}{{\ttfamily arXiv:2006.04822
  [hep-ph]}}.

\bibitem{Morel:2020dww}
L.~Morel, Z.~Yao, P.~Clad\'e, and S.~Guellati-Kh\'elifa, ``{Determination of
  the fine-structure constant with an accuracy of 81 parts per trillion},''
  \href{http://dx.doi.org/10.1038/s41586-020-2964-7}{{\em Nature} {\bfseries
  588} no.~7836, (2020) 61--65}.

\bibitem{Gerardin:2020gpp}
A.~G\'erardin, ``{The anomalous magnetic moment of the muon: status of Lattice
  QCD calculations},'' in {\em {38th International Symposium on Lattice Field
  Theory}}.
\newblock 12, 2020.
\newblock \href{http://arxiv.org/abs/2012.03931}{{\ttfamily arXiv:2012.03931
  [hep-lat]}}.

\bibitem{Lindner:2016bgg}
M.~Lindner, M.~Platscher, and F.~S. Queiroz, ``{A Call for New Physics : The
  Muon Anomalous Magnetic Moment and Lepton Flavor Violation},''
  \href{http://dx.doi.org/10.1016/j.physrep.2017.12.001}{{\em Phys. Rept.}
  {\bfseries 731} (2018) 1--82},
  \href{http://arxiv.org/abs/1610.06587}{{\ttfamily arXiv:1610.06587
  [hep-ph]}}.

\bibitem{Borsanyi:2020mff}
S.~Borsanyi {\em et~al.}, ``{Leading hadronic contribution to the muon 2
  magnetic moment from lattice QCD},''
  \href{http://arxiv.org/abs/2002.12347}{{\ttfamily arXiv:2002.12347
  [hep-lat]}}.

\bibitem{Passera:2008jk}
M.~Passera, W.~J. Marciano, and A.~Sirlin, ``{The Muon g-2 and the bounds on
  the Higgs boson mass},''
  \href{http://dx.doi.org/10.1103/PhysRevD.78.013009}{{\em Phys. Rev. D}
  {\bfseries 78} (2008) 013009},
  \href{http://arxiv.org/abs/0804.1142}{{\ttfamily arXiv:0804.1142 [hep-ph]}}.

\bibitem{Crivellin:2020zul}
A.~Crivellin, M.~Hoferichter, C.~A. Manzari, and M.~Montull, ``{Hadronic Vacuum
  Polarization: $(g-2)_\mu$ versus Global Electroweak Fits},''
  \href{http://dx.doi.org/10.1103/PhysRevLett.125.091801}{{\em Phys. Rev.
  Lett.} {\bfseries 125} no.~9, (2020) 091801},
  \href{http://arxiv.org/abs/2003.04886}{{\ttfamily arXiv:2003.04886
  [hep-ph]}}.

\bibitem{Keshavarzi:2020bfy}
A.~Keshavarzi, W.~J. Marciano, M.~Passera, and A.~Sirlin, ``{Muon $g-2$ and
  $\Delta \alpha$ connection},''
  \href{http://dx.doi.org/10.1103/PhysRevD.102.033002}{{\em Phys. Rev. D}
  {\bfseries 102} no.~3, (2020) 033002},
  \href{http://arxiv.org/abs/2006.12666}{{\ttfamily arXiv:2006.12666
  [hep-ph]}}.

\bibitem{Malaescu:2020zuc}
B.~Malaescu and M.~Schott, ``{Impact of correlations between $a_{\mu }$ and
  $\alpha _\text {QED}$ on the EW fit},''
  \href{http://dx.doi.org/10.1140/epjc/s10052-021-08848-9}{{\em Eur. Phys. J.
  C} {\bfseries 81} no.~1, (2021) 46},
  \href{http://arxiv.org/abs/2008.08107}{{\ttfamily arXiv:2008.08107
  [hep-ph]}}.

\bibitem{Giudice:2012ms}
G.~F. Giudice, P.~Paradisi, and M.~Passera, ``{Testing new physics with the
  electron g-2},'' \href{http://dx.doi.org/10.1007/JHEP11(2012)113}{{\em JHEP}
  {\bfseries 11} (2012) 113}, \href{http://arxiv.org/abs/1208.6583}{{\ttfamily
  arXiv:1208.6583 [hep-ph]}}.

\bibitem{Mohapatra:1986bd}
R.~N. Mohapatra and J.~W.~F. Valle, ``{Neutrino Mass and Baryon Number
  Nonconservation in Superstring Models},''
  \href{http://dx.doi.org/10.1103/PhysRevD.34.1642}{{\em Phys. Rev.} {\bfseries
  D34} (1986) 1642}.
[,235(1986)].

\bibitem{Abada:2007ux}
A.~Abada, C.~Biggio, F.~Bonnet, M.~B. Gavela, and T.~Hambye, ``{Low energy
  effects of neutrino masses},''
  \href{http://dx.doi.org/10.1088/1126-6708/2007/12/061}{{\em JHEP} {\bfseries
  12} (2007) 061},
\href{http://arxiv.org/abs/0707.4058}{{\ttfamily arXiv:0707.4058 [hep-ph]}}.

\bibitem{Gavela:2009cd}
M.~B. Gavela, T.~Hambye, D.~Hernandez, and P.~Hernandez, ``{Minimal Flavour
  Seesaw Models},'' \href{http://dx.doi.org/10.1088/1126-6708/2009/09/038}{{\em
  JHEP} {\bfseries 09} (2009) 038},
\href{http://arxiv.org/abs/0906.1461}{{\ttfamily arXiv:0906.1461 [hep-ph]}}.

\bibitem{Ibanez:2009du}
D.~Ibanez, S.~Morisi, and J.~W.~F. Valle, ``{Inverse tri-bimaximal type-III
  seesaw and lepton flavor violation},''
  \href{http://dx.doi.org/10.1103/PhysRevD.80.053015}{{\em Phys. Rev.}
  {\bfseries D80} (2009) 053015},
\href{http://arxiv.org/abs/0907.3109}{{\ttfamily arXiv:0907.3109 [hep-ph]}}.

\bibitem{Ma:2009kh}
E.~Ma, ``{Inverse Seesaw Neutrino Mass from Lepton Triplets in the
  $U(1)_\Sigma$ Model},''
  \href{http://dx.doi.org/10.1142/S0217732309031867}{{\em Mod. Phys. Lett.}
  {\bfseries A24} (2009) 2491--2495},
\href{http://arxiv.org/abs/0905.2972}{{\ttfamily arXiv:0905.2972 [hep-ph]}}.

\bibitem{Eboli:2011ia}
O.~J.~P. Eboli, J.~Gonzalez-Fraile, and M.~C. Gonzalez-Garcia, ``{Neutrino
  Masses at LHC: Minimal Lepton Flavour Violation in Type-III See-saw},''
  \href{http://dx.doi.org/10.1007/JHEP12(2011)009}{{\em JHEP} {\bfseries 12}
  (2011) 009},
\href{http://arxiv.org/abs/1108.0661}{{\ttfamily arXiv:1108.0661 [hep-ph]}}.

\bibitem{Morisi:2012hu}
S.~Morisi, E.~Peinado, and A.~Vicente, ``{Flavor origin of R-parity},''
  \href{http://dx.doi.org/10.1088/0954-3899/40/8/085004}{{\em J. Phys.}
  {\bfseries G40} (2013) 085004},
\href{http://arxiv.org/abs/1212.4145}{{\ttfamily arXiv:1212.4145 [hep-ph]}}.

\bibitem{Aguilar-Saavedra:2013twa}
J.~A. Aguilar-Saavedra, P.~M. Boavida, and F.~R. Joaquim, ``{Flavored searches
  for type-III seesaw mechanism at the LHC},''
  \href{http://dx.doi.org/10.1103/PhysRevD.88.113008}{{\em Phys. Rev.}
  {\bfseries D88} (2013) 113008},
\href{http://arxiv.org/abs/1308.3226}{{\ttfamily arXiv:1308.3226 [hep-ph]}}.

\bibitem{Law:2013gma}
S.~S.~C. Law and K.~L. McDonald, ``{Generalized inverse seesaw mechanisms},''
  \href{http://dx.doi.org/10.1103/PhysRevD.87.113003}{{\em Phys. Rev.}
  {\bfseries D87} no.~11, (2013) 113003},
\href{http://arxiv.org/abs/1303.4887}{{\ttfamily arXiv:1303.4887 [hep-ph]}}.

\bibitem{CentellesChulia:2020dfh}
S.~Centelles~Chuli\'a, R.~Srivastava, and A.~Vicente, ``{The inverse seesaw
  family: Dirac and Majorana},''
  \href{http://dx.doi.org/10.1007/JHEP03(2021)248}{{\em JHEP} {\bfseries 03}
  (2021) 248}, \href{http://arxiv.org/abs/2011.06609}{{\ttfamily
  arXiv:2011.06609 [hep-ph]}}.

\bibitem{Abada:2008ea}
A.~Abada, C.~Biggio, F.~Bonnet, M.~B. Gavela, and T.~Hambye, ``{$\mu \to e
  \gamma$ and $\tau \to l \gamma$ decays in the fermion triplet seesaw
  model},'' \href{http://dx.doi.org/10.1103/PhysRevD.78.033007}{{\em Phys.
  Rev.} {\bfseries D78} (2008) 033007},
\href{http://arxiv.org/abs/0803.0481}{{\ttfamily arXiv:0803.0481 [hep-ph]}}.

\bibitem{Franceschini:2008pz}
R.~Franceschini, T.~Hambye, and A.~Strumia, ``{Type-III see-saw at LHC},''
  \href{http://dx.doi.org/10.1103/PhysRevD.78.033002}{{\em Phys. Rev.}
  {\bfseries D78} (2008) 033002},
\href{http://arxiv.org/abs/0805.1613}{{\ttfamily arXiv:0805.1613 [hep-ph]}}.

\bibitem{Das:2020uer}
A.~Das and S.~Mandal, ``{Bounds on the triplet fermions in type-III seesaw and
  implications for collider searches},''
  \href{http://dx.doi.org/10.1016/j.nuclphysb.2021.115374}{{\em Nucl. Phys. B}
  {\bfseries 966} (2021) 115374},
  \href{http://arxiv.org/abs/2006.04123}{{\ttfamily arXiv:2006.04123
  [hep-ph]}}.

\bibitem{Ashanujjaman:2021jhi}
S.~Ashanujjaman and K.~Ghosh, ``{Type-III Seesaw: Phenomenological Implications
  of the Information Lost in Decoupling from High-Energy to Low-Energy},''
  \href{http://arxiv.org/abs/2102.09536}{{\ttfamily arXiv:2102.09536
  [hep-ph]}}.

\bibitem{Mandal:2021acg}
S.~Mandal, J.~C. Rom\~ao, R.~Srivastava, and J.~W.~F. Valle, ``{Dynamical
  inverse seesaw mechanism as a simple benchmark for electroweak breaking and
  Higgs boson studies},'' \href{http://arxiv.org/abs/2103.02670}{{\ttfamily
  arXiv:2103.02670 [hep-ph]}}.

\bibitem{Abada:2014kba}
A.~Abada, M.~E. Krauss, W.~Porod, F.~Staub, A.~Vicente, and C.~Weiland,
  ``{Lepton flavor violation in low-scale seesaw models: SUSY and non-SUSY
  contributions},'' \href{http://dx.doi.org/10.1007/JHEP11(2014)048}{{\em JHEP}
  {\bfseries 11} (2014) 048}, \href{http://arxiv.org/abs/1408.0138}{{\ttfamily
  arXiv:1408.0138 [hep-ph]}}.

\bibitem{Biggio:2019eeo}
C.~Biggio, E.~Fernandez-Martinez, M.~Filaci, J.~Hernandez-Garcia, and
  J.~Lopez-Pavon, ``{Global Bounds on the Type-III Seesaw},''
  \href{http://dx.doi.org/10.1007/JHEP05(2020)022}{{\em JHEP} {\bfseries 05}
  (2020) 022}, \href{http://arxiv.org/abs/1911.11790}{{\ttfamily
  arXiv:1911.11790 [hep-ph]}}.

\bibitem{Abdallah:2011ew}
W.~Abdallah, A.~Awad, S.~Khalil, and H.~Okada, ``{Muon Anomalous Magnetic
  Moment and $\mu \to e \gamma$ in $B-L$ Model with Inverse Seesaw},''
  \href{http://dx.doi.org/10.1140/epjc/s10052-012-2108-9}{{\em Eur. Phys. J. C}
  {\bfseries 72} (2012) 2108}, \href{http://arxiv.org/abs/1105.1047}{{\ttfamily
  arXiv:1105.1047 [hep-ph]}}.

\bibitem{Khalil:2015wua}
S.~Khalil and C.~S. Un, ``{Muon Anomalous Magnetic Moment in SUSY B-L Model
  with Inverse Seesaw},''
  \href{http://dx.doi.org/10.1016/j.physletb.2016.10.035}{{\em Phys. Lett. B}
  {\bfseries 763} (2016) 164--168},
  \href{http://arxiv.org/abs/1509.05391}{{\ttfamily arXiv:1509.05391
  [hep-ph]}}.

\bibitem{Cao:2019evo}
J.~Cao, J.~Lian, L.~Meng, Y.~Yue, and P.~Zhu, ``{Anomalous muon magnetic moment
  in the inverse seesaw extended next-to-minimal supersymmetric standard
  model},'' \href{http://dx.doi.org/10.1103/PhysRevD.101.095009}{{\em Phys.
  Rev. D} {\bfseries 101} no.~9, (2020) 095009},
  \href{http://arxiv.org/abs/1912.10225}{{\ttfamily arXiv:1912.10225
  [hep-ph]}}.

\bibitem{Dinh:2020pqn}
L.~T. Hue, P.~N. Thanh, and T.~D. Tham, ``{Anomalous Magnetic Dipole Moment
  \((g-2)\mu\) in 3-3-1 Model with Inverse Seesaw Neutrinos},''
  \href{http://dx.doi.org/10.15625/0868-3166/30/3/14963}{{\em Commun. in Phys.}
  {\bfseries 30} no.~3, (2020) 221--230}.

\bibitem{Cao:2021lmj}
J.~Cao, Y.~He, J.~Lian, D.~Zhang, and P.~Zhu, ``{Electron and Muon Anomalous
  Magnetic Moments in the Inverse Seesaw Extended NMSSM},''
  \href{http://arxiv.org/abs/2102.11355}{{\ttfamily arXiv:2102.11355
  [hep-ph]}}.

\bibitem{Nomura:2021adf}
T.~Nomura, H.~Okada, and P.~Sanyal, ``{A radiatively induced inverse seesaw
  model with hidden $U(1)$ gauge symmetry},''
  \href{http://arxiv.org/abs/2103.09494}{{\ttfamily arXiv:2103.09494
  [hep-ph]}}.

\bibitem{Mondal:2021vou}
T.~Mondal and H.~Okada, ``{Inverse seesaw and $(g-2)$ anomalies in $B-L$
  extended two Higgs doublet model},''
  \href{http://arxiv.org/abs/2103.13149}{{\ttfamily arXiv:2103.13149
  [hep-ph]}}.

\bibitem{Hue:2021xap}
L.~T. Hue, H.~T. Hung, N.~T. Tham, t.~H.~N. Long, and T.~Phong~Nguyen, ``{Large
  $(g-2)_{\mu}$ and signals of decays $e_b\rightarrow e_a\gamma$ in a 3-3-1
  model with inverse seesaw neutrinos},''
  \href{http://arxiv.org/abs/2104.01840}{{\ttfamily arXiv:2104.01840
  [hep-ph]}}.

\bibitem{CarcamoHernandez:2021iat}
A.~E. C\'arcamo~Hern\'andez, C.~Espinoza, J.~Carlos G\'omez-Izquierdo, and
  M.~Mondrag\'on, ``{Fermion masses and mixings, dark matter, leptogenesis and
  $g-2$ muon anomaly in an extended 2HDM with inverse seesaw},''
  \href{http://arxiv.org/abs/2104.02730}{{\ttfamily arXiv:2104.02730
  [hep-ph]}}.

\bibitem{Dermisek:2013gta}
R.~Dermisek and A.~Raval, ``{Explanation of the Muon g-2 Anomaly with
  Vectorlike Leptons and its Implications for Higgs Decays},''
  \href{http://dx.doi.org/10.1103/PhysRevD.88.013017}{{\em Phys. Rev. D}
  {\bfseries 88} (2013) 013017},
  \href{http://arxiv.org/abs/1305.3522}{{\ttfamily arXiv:1305.3522 [hep-ph]}}.

\bibitem{Poh:2017tfo}
Z.~Poh and S.~Raby, ``{Vectorlike leptons: Muon g-2 anomaly, lepton flavor
  violation, Higgs boson decays, and lepton nonuniversality},''
  \href{http://dx.doi.org/10.1103/PhysRevD.96.015032}{{\em Phys. Rev. D}
  {\bfseries 96} no.~1, (2017) 015032},
  \href{http://arxiv.org/abs/1705.07007}{{\ttfamily arXiv:1705.07007
  [hep-ph]}}.

\bibitem{Kowalska:2017iqv}
K.~Kowalska and E.~M. Sessolo, ``{Expectations for the muon g-2 in simplified
  models with dark matter},''
  \href{http://dx.doi.org/10.1007/JHEP09(2017)112}{{\em JHEP} {\bfseries 09}
  (2017) 112}, \href{http://arxiv.org/abs/1707.00753}{{\ttfamily
  arXiv:1707.00753 [hep-ph]}}.

\bibitem{Megias:2017dzd}
E.~Megias, M.~Quiros, and L.~Salas, ``{$g_\mu-2$ from Vector-Like Leptons in
  Warped Space},'' \href{http://dx.doi.org/10.1007/JHEP05(2017)016}{{\em JHEP}
  {\bfseries 05} (2017) 016}, \href{http://arxiv.org/abs/1701.05072}{{\ttfamily
  arXiv:1701.05072 [hep-ph]}}.

\bibitem{Chiang:2017tai}
C.-W. Chiang, H.~Okada, and E.~Senaha, ``{Dark matter, muon $g-2$, electric
  dipole moments, and $Z\to \ell_i^+ \ell_j^-$ in a one-loop induced neutrino
  model},'' \href{http://dx.doi.org/10.1103/PhysRevD.96.015002}{{\em Phys. Rev.
  D} {\bfseries 96} no.~1, (2017) 015002},
  \href{http://arxiv.org/abs/1703.09153}{{\ttfamily arXiv:1703.09153
  [hep-ph]}}.

\bibitem{Calibbi:2018rzv}
L.~Calibbi, R.~Ziegler, and J.~Zupan, ``{Minimal models for dark matter and the
  muon g\ensuremath{-}2 anomaly},''
  \href{http://dx.doi.org/10.1007/JHEP07(2018)046}{{\em JHEP} {\bfseries 07}
  (2018) 046}, \href{http://arxiv.org/abs/1804.00009}{{\ttfamily
  arXiv:1804.00009 [hep-ph]}}.

\bibitem{Arnan:2019uhr}
P.~Arnan, A.~Crivellin, M.~Fedele, and F.~Mescia, ``{Generic Loop Effects of
  New Scalars and Fermions in $b\to s\ell^+\ell^-$, $(g-2)_\mu$ and a
  Vector-like $4^{\rm th}$ Generation},''
  \href{http://dx.doi.org/10.1007/JHEP06(2019)118}{{\em JHEP} {\bfseries 06}
  (2019) 118}, \href{http://arxiv.org/abs/1904.05890}{{\ttfamily
  arXiv:1904.05890 [hep-ph]}}.

\bibitem{Kawamura:2019rth}
J.~Kawamura, S.~Raby, and A.~Trautner, ``{Complete vectorlike fourth family and
  new U(1)' for muon anomalies},''
  \href{http://dx.doi.org/10.1103/PhysRevD.100.055030}{{\em Phys. Rev. D}
  {\bfseries 100} no.~5, (2019) 055030},
  \href{http://arxiv.org/abs/1906.11297}{{\ttfamily arXiv:1906.11297
  [hep-ph]}}.

\bibitem{Calibbi:2020emz}
L.~Calibbi, M.~L. L\'opez-Ib\'a\~nez, A.~Melis, and O.~Vives, ``{Muon and
  electron $g-2$ and lepton masses in flavor models},''
  \href{http://dx.doi.org/10.1007/JHEP06(2020)087}{{\em JHEP} {\bfseries 06}
  (2020) 087}, \href{http://arxiv.org/abs/2003.06633}{{\ttfamily
  arXiv:2003.06633 [hep-ph]}}.

\bibitem{Frank:2020smf}
M.~Frank and I.~Saha, ``{Muon anomalous magnetic moment in two-Higgs-doublet
  models with vectorlike leptons},''
  \href{http://dx.doi.org/10.1103/PhysRevD.102.115034}{{\em Phys. Rev. D}
  {\bfseries 102} no.~11, (2020) 115034},
  \href{http://arxiv.org/abs/2008.11909}{{\ttfamily arXiv:2008.11909
  [hep-ph]}}.

\bibitem{Chun:2020uzw}
E.~J. Chun and T.~Mondal, ``{Explaining $g-2$ anomalies in two Higgs doublet
  model with vector-like leptons},''
  \href{http://dx.doi.org/10.1007/JHEP11(2020)077}{{\em JHEP} {\bfseries 11}
  (2020) 077}, \href{http://arxiv.org/abs/2009.08314}{{\ttfamily
  arXiv:2009.08314 [hep-ph]}}.

\bibitem{Chakrabarty:2020jro}
N.~Chakrabarty, ``{Doubly charged scalars and vector-like leptons confronting
  the muon g-2 anomaly and Higgs vacuum stability},''
  \href{http://arxiv.org/abs/2010.05215}{{\ttfamily arXiv:2010.05215
  [hep-ph]}}.

\bibitem{Chen:2020tfr}
K.-F. Chen, C.-W. Chiang, and K.~Yagyu, ``{An explanation for the muon and
  electron $g-2$ anomalies and dark matter},''
  \href{http://dx.doi.org/10.1007/JHEP09(2020)119}{{\em JHEP} {\bfseries 09}
  (2020) 119}, \href{http://arxiv.org/abs/2006.07929}{{\ttfamily
  arXiv:2006.07929 [hep-ph]}}.

\bibitem{Jana:2020joi}
S.~Jana, P.~K. Vishnu, W.~Rodejohann, and S.~Saad, ``{Dark matter assisted
  lepton anomalous magnetic moments and neutrino masses},''
  \href{http://dx.doi.org/10.1103/PhysRevD.102.075003}{{\em Phys. Rev. D}
  {\bfseries 102} no.~7, (2020) 075003},
  \href{http://arxiv.org/abs/2008.02377}{{\ttfamily arXiv:2008.02377
  [hep-ph]}}.

\bibitem{Dermisek:2020cod}
R.~Dermisek, K.~Hermanek, and N.~McGinnis, ``{Highly enhanced contributions of
  heavy Higgs bosons and new leptons to muon $g-2$ and other observables},''
  \href{http://arxiv.org/abs/2011.11812}{{\ttfamily arXiv:2011.11812
  [hep-ph]}}.

\bibitem{Das:2020hpd}
P.~Das, M.~K. Das, and N.~Khan, ``{A new feasible dark matter region in the
  singlet scalar scotogenic model},''
  \href{http://dx.doi.org/10.1016/j.nuclphysb.2021.115307}{{\em Nucl. Phys. B}
  {\bfseries 964} (2021) 115307},
  \href{http://arxiv.org/abs/2001.04070}{{\ttfamily arXiv:2001.04070
  [hep-ph]}}.

\bibitem{Baker:2021yli}
M.~J. Baker, P.~Cox, and R.~R. Volkas, ``{Radiative Muon Mass Models and
  $(g-2)_\mu$},'' \href{http://arxiv.org/abs/2103.13401}{{\ttfamily
  arXiv:2103.13401 [hep-ph]}}.

\bibitem{Dermisek:2021ajd}
R.~Dermisek, K.~Hermanek, and N.~McGinnis, ``{Muon $g-2$ in two Higgs doublet
  models with vectorlike leptons},''
  \href{http://arxiv.org/abs/2103.05645}{{\ttfamily arXiv:2103.05645
  [hep-ph]}}.

\bibitem{Das:2021zea}
P.~Das, M.~Kumar~Das, and N.~Khan, ``{The FIMP-WIMP dark matter and Muon g-2 in
  the extended singlet scalar model},''
  \href{http://arxiv.org/abs/2104.03271}{{\ttfamily arXiv:2104.03271
  [hep-ph]}}.

\bibitem{Chiang:2021pma}
C.-W. Chiang and K.~Yagyu, ``{Radiative Seesaw Mechanism for Charged
  Leptons},'' \href{http://arxiv.org/abs/2104.00890}{{\ttfamily
  arXiv:2104.00890 [hep-ph]}}.

\bibitem{Arcadi:2021cwg}
G.~Arcadi, L.~Calibbi, M.~Fedele, and F.~Mescia, ``{Muon $g-2$ and
  $B$-anomalies from Dark Matter},''
  \href{http://arxiv.org/abs/2104.03228}{{\ttfamily arXiv:2104.03228
  [hep-ph]}}.

\bibitem{Arbelaez:2020rbq}
C.~Arbel\'aez, R.~Cepedello, R.~M. Fonseca, and M.~Hirsch, ``{$(g-2)$ anomalies
  and neutrino mass},''
  \href{http://dx.doi.org/10.1103/PhysRevD.102.075005}{{\em Phys. Rev. D}
  {\bfseries 102} no.~7, (2020) 075005},
  \href{http://arxiv.org/abs/2007.11007}{{\ttfamily arXiv:2007.11007
  [hep-ph]}}.

\bibitem{Buttazzo:2020eyl}
D.~Buttazzo and P.~Paradisi, ``{Probing the muon g-2 anomaly at a Muon
  Collider},'' \href{http://arxiv.org/abs/2012.02769}{{\ttfamily
  arXiv:2012.02769 [hep-ph]}}.

\bibitem{Yin:2020afe}
W.~Yin and M.~Yamaguchi, ``{Muon $g-2$ at multi-TeV muon collider},''
  \href{http://arxiv.org/abs/2012.03928}{{\ttfamily arXiv:2012.03928
  [hep-ph]}}.

\bibitem{Capdevilla:2021rwo}
R.~Capdevilla, D.~Curtin, Y.~Kahn, and G.~Krnjaic, ``{A No-Lose Theorem for
  Discovering the New Physics of $(g-2)_\mu$ at Muon Colliders},''
  \href{http://arxiv.org/abs/2101.10334}{{\ttfamily arXiv:2101.10334
  [hep-ph]}}.

\bibitem{tHooft:1979rat}
G.~'t~Hooft, ``{Naturalness, chiral symmetry, and spontaneous chiral symmetry
  breaking},'' \href{http://dx.doi.org/10.1007/978-1-4684-7571-5_9}{{\em NATO
  Sci. Ser. B} {\bfseries 59} (1980) 135--157}.

\bibitem{Crivellin:2018qmi}
A.~Crivellin, M.~Hoferichter, and P.~Schmidt-Wellenburg, ``{Combined
  explanations of $(g-2)_{\mu,e}$ and implications for a large muon EDM},''
  \href{http://dx.doi.org/10.1103/PhysRevD.98.113002}{{\em Phys. Rev. D}
  {\bfseries 98} no.~11, (2018) 113002},
  \href{http://arxiv.org/abs/1807.11484}{{\ttfamily arXiv:1807.11484
  [hep-ph]}}.

\bibitem{Romao:2012pq}
J.~C. Romao and J.~P. Silva, ``{A resource for signs and Feynman diagrams of
  the Standard Model},''
  \href{http://dx.doi.org/10.1142/S0217751X12300256}{{\em Int. J. Mod. Phys. A}
  {\bfseries 27} (2012) 1230025},
  \href{http://arxiv.org/abs/1209.6213}{{\ttfamily arXiv:1209.6213 [hep-ph]}}.

\bibitem{Patel:2015tea}
H.~H. Patel, ``{Package-X: A Mathematica package for the analytic calculation
  of one-loop integrals},''
  \href{http://dx.doi.org/10.1016/j.cpc.2015.08.017}{{\em Comput. Phys.
  Commun.} {\bfseries 197} (2015) 276--290},
  \href{http://arxiv.org/abs/1503.01469}{{\ttfamily arXiv:1503.01469
  [hep-ph]}}.

\bibitem{Barr:1990vd}
S.~M. Barr and A.~Zee, ``{Electric Dipole Moment of the Electron and of the
  Neutron},'' \href{http://dx.doi.org/10.1103/PhysRevLett.65.21}{{\em Phys.
  Rev. Lett.} {\bfseries 65} (1990) 21--24}. [Erratum: Phys.Rev.Lett. 65, 2920
  (1990)].

\bibitem{deSalas:2020pgw}
P.~F. de~Salas, D.~V. Forero, S.~Gariazzo, P.~Mart\'\i{}nez-Mirav\'e, O.~Mena,
  C.~A. Ternes, M.~T\'ortola, and J.~W.~F. Valle, ``{2020 global reassessment
  of the neutrino oscillation picture},''
  \href{http://dx.doi.org/10.1007/JHEP02(2021)071}{{\em JHEP} {\bfseries 02}
  (2021) 071}, \href{http://arxiv.org/abs/2006.11237}{{\ttfamily
  arXiv:2006.11237 [hep-ph]}}.

\bibitem{Cordero-Carrion:2018xre}
I.~Cordero-Carri\'on, M.~Hirsch, and A.~Vicente, ``{Master Majorana neutrino
  mass parametrization},''
  \href{http://dx.doi.org/10.1103/PhysRevD.99.075019}{{\em Phys. Rev. D}
  {\bfseries 99} no.~7, (2019) 075019},
  \href{http://arxiv.org/abs/1812.03896}{{\ttfamily arXiv:1812.03896
  [hep-ph]}}.

\bibitem{Cordero-Carrion:2019qtu}
I.~Cordero-Carri\'on, M.~Hirsch, and A.~Vicente, ``{General parametrization of
  Majorana neutrino mass models},''
  \href{http://dx.doi.org/10.1103/PhysRevD.101.075032}{{\em Phys. Rev. D}
  {\bfseries 101} no.~7, (2020) 075032},
  \href{http://arxiv.org/abs/1912.08858}{{\ttfamily arXiv:1912.08858
  [hep-ph]}}.

\bibitem{Casas:2001sr}
J.~A. Casas and A.~Ibarra, ``{Oscillating neutrinos and $\mu \to e, \gamma$},''
  \href{http://dx.doi.org/10.1016/S0550-3213(01)00475-8}{{\em Nucl. Phys. B}
  {\bfseries 618} (2001) 171--204},
  \href{http://arxiv.org/abs/hep-ph/0103065}{{\ttfamily arXiv:hep-ph/0103065}}.

\bibitem{Crivellin:2021rbq}
A.~Crivellin and M.~Hoferichter, ``{Consequences of chirally enhanced
  explanations of $(g-2)_\mu$ for $h\to \mu\mu$ and $Z\to \mu\mu$},''
  \href{http://arxiv.org/abs/2104.03202}{{\ttfamily arXiv:2104.03202
  [hep-ph]}}.

\bibitem{Zyla:2020zbs}
{\bfseries Particle Data Group} Collaboration, P.~A. Zyla {\em et~al.},
  ``{Review of Particle Physics},''
  \href{http://dx.doi.org/10.1093/ptep/ptaa104}{{\em PTEP} {\bfseries 2020}
  no.~8, (2020) 083C01}.

\bibitem{Sirunyan:2020two}
{\bfseries CMS} Collaboration, A.~M. Sirunyan {\em et~al.}, ``{Evidence for
  Higgs boson decay to a pair of muons},''
  \href{http://dx.doi.org/10.1007/JHEP01(2021)148}{{\em JHEP} {\bfseries 01}
  (2021) 148}, \href{http://arxiv.org/abs/2009.04363}{{\ttfamily
  arXiv:2009.04363 [hep-ex]}}.

\bibitem{Fajfer:2021cxa}
S.~Fajfer, J.~F. Kamenik, and M.~Tammaro, ``{Interplay of New Physics Effects
  in $(g-2)_\ell$ and $h\to\ell^+\ell^-$ -- Lessons from SMEFT},''
  \href{http://arxiv.org/abs/2103.10859}{{\ttfamily arXiv:2103.10859
  [hep-ph]}}.

\bibitem{Aad:2020fzq}
{\bfseries ATLAS} Collaboration, G.~Aad {\em et~al.}, ``{Search for type-III
  seesaw heavy leptons in dilepton final states in $pp$ collisions at
  $\sqrt{s}$ = 13 TeV with the ATLAS detector},'' {\em Eur. Phys. J. C}
  {\bfseries 81} no.~3, (2021) 218,
  \href{http://arxiv.org/abs/2008.07949}{{\ttfamily arXiv:2008.07949
  [hep-ex]}}.

\bibitem{CMS:2012ra}
{\bfseries CMS} Collaboration, S.~Chatrchyan {\em et~al.}, ``{Search for Heavy
  Lepton Partners of Neutrinos in Proton-Proton Collisions in the Context of
  the Type III Seesaw Mechanism},''
  \href{http://dx.doi.org/10.1016/j.physletb.2012.10.070}{{\em Phys. Lett. B}
  {\bfseries 718} (2012) 348--368},
  \href{http://arxiv.org/abs/1210.1797}{{\ttfamily arXiv:1210.1797 [hep-ex]}}.

\bibitem{Sirunyan:2017qkz}
{\bfseries CMS} Collaboration, A.~M. Sirunyan {\em et~al.}, ``{Search for
  Evidence of the Type-III Seesaw Mechanism in Multilepton Final States in
  Proton-Proton Collisions at $\sqrt{s}=13\text{ }\text{ }\mathrm{TeV}$},''
  \href{http://dx.doi.org/10.1103/PhysRevLett.119.221802}{{\em Phys. Rev.
  Lett.} {\bfseries 119} no.~22, (2017) 221802},
  \href{http://arxiv.org/abs/1708.07962}{{\ttfamily arXiv:1708.07962
  [hep-ex]}}.

\bibitem{Biggio:2011ja}
C.~Biggio and F.~Bonnet, ``{Implementation of the Type III Seesaw Model in
  FeynRules/MadGraph and Prospects for Discovery with Early LHC Data},''
  \href{http://dx.doi.org/10.1140/epjc/s10052-012-1899-z}{{\em Eur. Phys. J. C}
  {\bfseries 72} (2012) 1899}, \href{http://arxiv.org/abs/1107.3463}{{\ttfamily
  arXiv:1107.3463 [hep-ph]}}.

\bibitem{Falkowski:2013jya}
A.~Falkowski, D.~M. Straub, and A.~Vicente, ``{Vector-like leptons: Higgs
  decays and collider phenomenology},''
  \href{http://dx.doi.org/10.1007/JHEP05(2014)092}{{\em JHEP} {\bfseries 05}
  (2014) 092}, \href{http://arxiv.org/abs/1312.5329}{{\ttfamily arXiv:1312.5329
  [hep-ph]}}.

\bibitem{Freitas:2020ttd}
F.~F. Freitas, J.~a. Gon\c{c}alves, A.~P. Morais, and R.~Pasechnik,
  ``{Phenomenology of vector-like leptons with Deep Learning at the Large
  Hadron Collider},'' \href{http://dx.doi.org/10.1007/JHEP01(2021)076}{{\em
  JHEP} {\bfseries 01} (2021) 076},
  \href{http://arxiv.org/abs/2010.01307}{{\ttfamily arXiv:2010.01307
  [hep-ph]}}.

\bibitem{Biggio:2008in}
C.~Biggio, ``{The Contribution of fermionic seesaws to the anomalous magnetic
  moment of leptons},''
  \href{http://dx.doi.org/10.1016/j.physletb.2008.09.004}{{\em Phys. Lett. B}
  {\bfseries 668} (2008) 378--384},
  \href{http://arxiv.org/abs/0806.2558}{{\ttfamily arXiv:0806.2558 [hep-ph]}}.

\bibitem{Chao:2008iw}
W.~Chao, ``{The Muon Magnetic Moment in the TeV Scale Seesaw Models},''
  \href{http://arxiv.org/abs/0806.0889}{{\ttfamily arXiv:0806.0889 [hep-ph]}}.

\bibitem{Chua:2010me}
C.-K. Chua and S.~S.~C. Law, ``{Phenomenological constraints on minimally
  coupled exotic lepton triplets},''
  \href{http://dx.doi.org/10.1103/PhysRevD.83.055010}{{\em Phys. Rev. D}
  {\bfseries 83} (2011) 055010},
  \href{http://arxiv.org/abs/1011.4730}{{\ttfamily arXiv:1011.4730 [hep-ph]}}.

\bibitem{Freitas:2014pua}
A.~Freitas, J.~Lykken, S.~Kell, and S.~Westhoff, ``{Testing the Muon g-2
  Anomaly at the LHC},'' \href{http://dx.doi.org/10.1007/JHEP09(2014)155}{{\em
  JHEP} {\bfseries 05} (2014) 145},
  \href{http://arxiv.org/abs/1402.7065}{{\ttfamily arXiv:1402.7065 [hep-ph]}}.
  [Erratum: JHEP 09, 155 (2014)].

\bibitem{Escribano:2020wua}
P.~Escribano and A.~Vicente, ``{Ultralight scalars in leptonic observables},''
  \href{http://dx.doi.org/10.1007/JHEP03(2021)240}{{\em JHEP} {\bfseries 03}
  (2021) 240}, \href{http://arxiv.org/abs/2008.01099}{{\ttfamily
  arXiv:2008.01099 [hep-ph]}}.

\bibitem{Staub:2008uz}
F.~Staub, ``{SARAH},''
\href{http://arxiv.org/abs/0806.0538}{{\ttfamily arXiv:0806.0538 [hep-ph]}}.

\bibitem{Staub:2009bi}
F.~Staub, ``{From Superpotential to Model Files for FeynArts and
  CalcHep/CompHep},'' \href{http://dx.doi.org/10.1016/j.cpc.2010.01.011}{{\em
  Comput. Phys. Commun.} {\bfseries 181} (2010) 1077--1086},
\href{http://arxiv.org/abs/0909.2863}{{\ttfamily arXiv:0909.2863 [hep-ph]}}.

\bibitem{Staub:2010jh}
F.~Staub, ``{Automatic Calculation of supersymmetric Renormalization Group
  Equations and Self Energies},''
  \href{http://dx.doi.org/10.1016/j.cpc.2010.11.030}{{\em Comput. Phys.
  Commun.} {\bfseries 182} (2011) 808--833},
\href{http://arxiv.org/abs/1002.0840}{{\ttfamily arXiv:1002.0840 [hep-ph]}}.

\bibitem{Staub:2012pb}
F.~Staub, ``{SARAH 3.2: Dirac Gauginos, UFO output, and more},''
  \href{http://dx.doi.org/10.1016/j.cpc.2013.02.019}{{\em Comput. Phys.
  Commun.} {\bfseries 184} (2013) 1792--1809},
\href{http://arxiv.org/abs/1207.0906}{{\ttfamily arXiv:1207.0906 [hep-ph]}}.

\bibitem{Staub:2013tta}
F.~Staub, ``{SARAH 4 : A tool for (not only SUSY) model builders},''
  \href{http://dx.doi.org/10.1016/j.cpc.2014.02.018}{{\em Comput. Phys.
  Commun.} {\bfseries 185} (2014) 1773--1790},
\href{http://arxiv.org/abs/1309.7223}{{\ttfamily arXiv:1309.7223 [hep-ph]}}.

\bibitem{Vicente:2015zba}
A.~Vicente, ``{Computer tools in particle physics},''
\href{http://arxiv.org/abs/1507.06349}{{\ttfamily arXiv:1507.06349 [hep-ph]}}.

\end{thebibliography}
\end{document}